\documentclass[aps,twocolumn,pre,floatfix]{revtex4-2}

\usepackage{xcolor,hyperref}
\hypersetup{
   colorlinks,
   linkcolor={blue!50!black},
   citecolor={blue!50!black},
   urlcolor={blue!80!black}
}

\usepackage{epsfig,amsmath}
\usepackage{bm}
\usepackage{soul,ulem}
\usepackage{hyperref}
\usepackage{footmisc}
\usepackage{wrapfig}
\DeclareMathAlphabet{\mathitbf}{OML}{cmm}{b}{it}
\newcommand{\pa}{\partial}

\renewcommand{\=}{\!=\!}

\definecolor{orange}{RGB}{255,127,0}
\definecolor{navyBlue}{RGB}{0,0,180}
\definecolor{magenta}{RGB}{255,0,255}
\definecolor{lightYellow}{RGB}{255,255,100}
\definecolor{brown}{RGB}{161,85,4}
\definecolor{violet}{RGB}{85,4,161}
\definecolor{darkGreen}{RGB}{4,161,85}
\definecolor{darkGreen2}{RGB}{0,100,0}
\definecolor{lightGray}{RGB}{180,180,180}

\DeclareMathAlphabet\mathbfcal{OMS}{cmsy}{b}{n}
\begin{document}

\title{Predicting activation-barrier and plasticity-onset statistics in a model of glasses}
\author{Makoto Suda$^1$}
\author{Edan Lerner$^2$}
\author{Eran Bouchbinder$^3$}
\affiliation{$^1$Department of Earth Science, Tohoku University, Sendai, Miyagi 980-8578, Japan\\
$^2$Institute for Theoretical Physics, University of Amsterdam, Science Park 904, 1098 XH Amsterdam, the Netherlands\\
$^3$Chemical and Biological Physics Department, Weizmann Institute of Science, Rehovot 7610001, Israel}

\begin{abstract}
A recently introduced anharmonic mean-field model unifiedly reproduced a broad range of low-temperature glass phenomena --- including harmonic nonphononic spectral properties, linear micromechanics and strongly driven elasto-plastic dynamics --- indicating that its underlying energy landscape is intrinsically glassy. Here, we apply a nonlinear modes framework to the model and derive analytic predictions for the asymptotic distributions of activation barriers $p(\Delta{U})\!\sim\!(\Delta{U})^{1/4}$ and the external force needed for the onset of plasticity $p(f_{\rm c})\sim f_{\rm c}^{2/3}$, for their extreme-value scaling and for $\langle\Delta{U}\rangle$ beyond the asymptotic regime. These predictions are expected to equally apply to the mean-field model and to finite-dimensional glasses. We develop efficient algorithms for sampling minima and saddles of the model's glassy potential energy landscape, and quantitatively confirm the theoretical predictions. This progress is enabled by identifying a subset of collective degrees of freedom that are physically relevant for activated glassy dynamics, which like the theoretical predictions should apply to realistic glasses.
\end{abstract}

\maketitle

\section{B\lowercase{ackground and motivation}}
Understanding the glassy phase of matter remains a major challenge. The need to develop simple models that capture the essence of ``glassiness'' and predict in a unified manner a broad range of glassy phenomena resulted in extensive efforts over the years~\cite{Kuhn1997-bl,Gurarie2003-qi,Gurevich2003-oo,Gurevich2005-ru,Parshin2007-od,Ritort2003-yu,Cavagna2009-wq,DeGiuli2014-oi,Franz2015-cc,Ros2019-nl,Folena2020-ud,Franz2025-tp,Vogel2026-xe}. Recent progress in computer glasses has given rise to important ``minimal selection criteria'' of models based on their vibrational density of states (VDoS) and the localization properties of vibrational modes. It has been shown that glasses feature in addition to extended phonons, which exist in both ordered/crystalline and disordered/noncrystalline solids~\cite{Kittel2004,Sette1998-gc,Grigera2003-ym,Parisi2003-ex,Leonforte2005-nm}, also quasi-localized nonphononic modes at the low-frequency end of the spectrum~\cite{Schober1991-in}. These nonphononic modes have been shown to follow a non-Debye VDoS ${\cal D}_{\!_{\rm G}\!}(\omega)$, where $\omega$ is the vibrational frequency, with universal properties~\cite{Baity-Jesi2015-pz,Lerner2016-ra,Mizuno2017-sc,Angelani2018-vt,Wang2019-jq,Bonfanti2020-nv,Lerner2021-en,Wang2025-so}.

It has been shown that ${\cal D}_{\!_{\rm G}\!}(\omega)$ features a pseudogapped quartic tail $\sim\!\omega^4$ in the $\omega\!\to\!0$ limit and that the nonphononic modes populating this tail are localized, i.e., reveal an ${\cal O}(1/N)$ participation ratio. Moreover, it has been suggested that at somewhat higher frequencies, ${\cal D}_{\!_{\rm G}\!}(\omega)$ features an intrinsic peak that is related to the conventional boson peak~\cite{Gurarie2003-qi,Gurevich2003-oo,Gurevich2005-ru,Parshin2007-od,Ramos2022-nf,Buchenau1984-dy,Malinovsky1986-vb,Schirmacher1998-fa,Nakayama2002-ve,Grigera2003-ym,Parisi2003-ex,Leonforte2005-nm,Chumakov2004-gb,Gonzalez-Jimenez2023-un,Festi2026-zu,Lerner2023-dp,Moriel2024-wp}, where harmonic vibrations are composed of many coupled quasi-localized modes (that are also hybridized with phonons, which is always the case in the thermodynamic limit, at any $\omega$)~\cite{Lerner2023-dp,Moriel2024-wp}. In terms of nonphononic degrees of freedom, the latter observations suggest that nonphononic vibrations become delocalized in the vicinity of the intrinsic boson peak frequency $\omega_{\rm p}$.

A recently introduced Hamiltonian mean-field model satisfied these criteria~\cite{Bouchbinder2021-dh,Rainone2021-am}. It has been subsequently shown to predict, without additional ad-hoc ingredients and/or parameters, linear micromechanical response properties that closely mirror their computer glass counterparts~\cite{Suda2026-cl}. Moreover, it has been shown that when externally driven well beyond linear response, deep into the strongly nonlinear regime, the mean-field model predicts essentially all the salient features of the oscillatory yielding phase transition and of mechanical memory formation observed in computer and laboratory glasses~\cite{Suda2025-kl}. These findings highlight the unifying nature of the model and its underlying glassy energy landscape, and call for further testing its predictive powers.

The first two aforementioned classes of glass phenomena, i.e., vibrational modes/spectra and linear micromechanical response, correspond to the harmonic regime derived from the model's Hessian. The last class, concerning elasto-plastic yielding transitions and memory formation, involves strongly anharmonic and nonlinear effects. Here, we apply to the model a nonlinear modes framework, originally developed in computer glasses~\cite{Gartner2016-vw,Gartner2016-lc}, to probe the anharmonic properties of the underlying energy landscape. We use the framework to derive analytic predictions for the landscape's activation barriers $\Delta{U}$ and the external force $f_{\rm c}$ needed to trigger plasticity. These predictions are expected to be equally valid for finite-dimensional glasses. Here, building on recent progress in understanding the model, we numerically explore its glassy landscape and quantitatively confirm the predictions.

\section{M\lowercase{odel formulation and a nonlinear modes framework}}
The model is defined by the following Hamiltonian of $N$ continuous, scalar degrees of freedom $\{x_i\}$,
\begin{equation}
    H\!=\!\frac{1}{2}\sum_{i=1}^{N} k_i x_i^2 + \frac{1}{4!}\sum_{i=1}^{N} x_i^4 + \sum_{i<j}J_{ij}x_ix_j -f(t)\sum_{i=1}^{N} x_i\,.
    \label{eq:Hamiltonian}
\end{equation}
Each $x_i$ is subjected to a local (``on-site'') potential, featuring both a random quadratic contribution, with a stiffness $k_i$ drawn uniformly over $[0,1]$, and a deterministic quartic contribution. In the absence of interaction, $\{x_i\}$ are displacements of anharmonic oscillators~\cite{Kuhn1997-bl,Gurevich2003-oo}.

Each oscillator interacts with all others bi-linearly, with disordered Gaussian coupling constants $J_{ij}$ of zero mean and variance $J^2\!/N$. The oscillators can also be subjected to a uniform, time-dependent external force $f(t)$, which is turned on after $H$ is settled into one of its minima under $f\=0$, a process that mimics a quench from a high temperature and involves displacing the oscillators from their initial $x_i\=0$ positions. Importantly, $H$ in Eq.~\eqref{eq:Hamiltonian} lacks translational invariance and hence phononic modes are a priori excluded.

The statistical properties of the minima of $H$ have been extensively studied at the harmonic level~\cite{Bouchbinder2021-dh,Rainone2021-am,Moriel2024-wp,Suda2026-cl}; see the summary above and additional discussion below. Yet, we first focus on a nonlinear modes framework that allows one to probe anharmonic properties of the model's energy landscape. We are directly inspired by the developments in finite-dimensional computer glasses. The framework was originally introduced in~\cite{Gartner2016-vw,Gartner2016-lc}, and subsequently it was shown to robustly identify soft degrees of freedom underlying plastic rearrangements~\cite{Lerner2016-ep,Kapteijns2020-fu,Richard2020-vr}.

For a generalized $N$-dimensional direction ${\bm z}$, defined at a minimum ${\bm x}_0$ of $H$, consider a one-dimensional energy profile $\delta H(s)\!\equiv\!H({\bm x}_0 + s\,{\bm z})\!-\!H({\bm x}_0)$. Here, the scalar $s$ parametrizes the magnitude of the displacement along ${\bm z}$ in the anharmonic energy landscape. Since $H$ is quartic, we obtain an exact expansion
\begin{equation}
    \delta{H}(s) =b_2({\bm z})s^2/2+b_3({\bm z})s^3/3!+b_4({\bm z})s^4/4!\,,
\label{eq:energy_expansion}
\end{equation}
where the directional expansion coefficients $b_{2,3,4}$ are given in Appendix~\ref{sec_activation_barrier:appendix_A}. Since the overall sign of ${\bm z}$ is of no importance, specifically note that $b_3(-{\bm z})\=-b_3({\bm z})$, we hereafter fix the sign of ${\bm z}$ such that $b_3\!<\!0$. Probing all possible landscape directions is not practical in general and not physically insightful/relevant. Rather, one is interested in selecting a subset of directions that are more physically relevant, most notably for activated dynamics.

To this aim, we define a normalized displacement $y\!\equiv\!s/(b_2/|b_3|)$ such that Eq.~\eqref{eq:energy_expansion} becomes $\delta{H}(y)\= \mathcal{E}\, v(y;\eta)$, with $v(y;\eta)\!\equiv\!y^2\!/2-y^3\!/3!+\eta^{-1}y^4\!/4!$, and the combination of expansion coefficients $\mathcal{E}\!\equiv\!b_2^3/b_3^2$ and $\eta\!\equiv\!b_3^2/b_2b_4$. In this form, $\eta$ controls the shape of the energy profile, e.g., whether it features a single minimum or two, while $\mathcal{E}$ determines the characteristic energy scale. Specifically, when $\delta{H}(y)$ features two minima, which is generic in glassy landscapes~\cite{Anderson1972-bi,Phillips1987-ti,Heuer2008-ou,Falk2011-nd,Berthier2023-cf,Blaber2026-ic}, $\mathcal{E}$ sets the scale of the barrier separating them. Moreover, since $b_2({\bm z})$ is the stiffness of ${\bm z}$ (i.e., the curvature of the landscape along ${\bm z}$) and $b_3({\bm z})$ is the asymmetry of the landscape basin in this direction, low barriers correspond to soft (small $b_2$) and asymmetric (large $|b_3|$) landscape directions. Consequently, and following~\cite{Kapteijns2020-fu}, we focus on directions that minimize
\begin{equation}
    \mathcal{E}({\bm z}) \equiv [b_2({\bm z})]^3/[b_3({\bm z})]^2\,.
\label{eq:cost_function}
\end{equation}

\begin{figure}[t!]
\centering
\includegraphics[width=0.985\columnwidth]{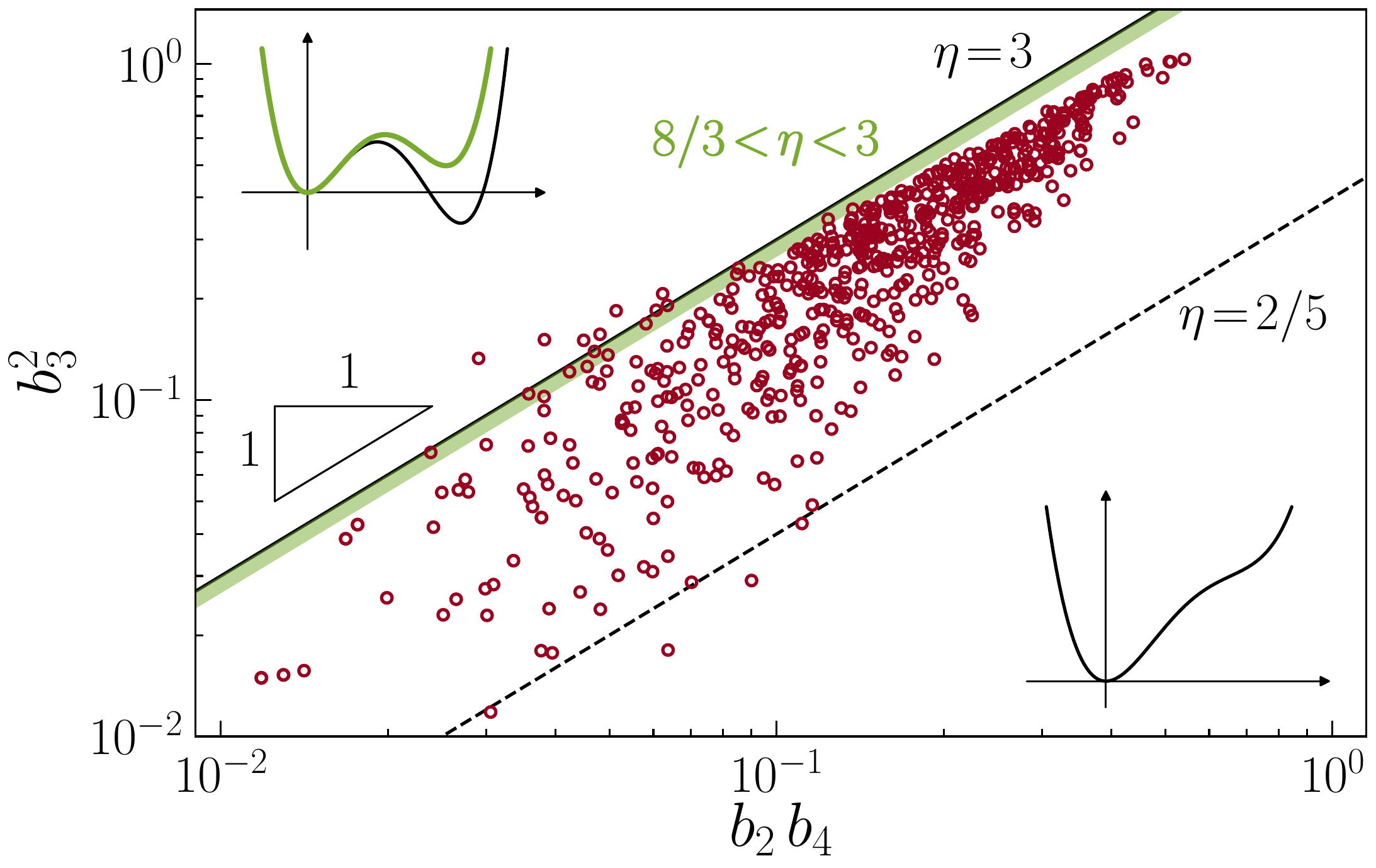}
\caption{The expansion coefficients $b_3^2$ vs~$b_2b_4$, evaluated along distinct cubic modes in a single quenched-disorder realization at $N\!=\!8192$ and $J\!=\!0.5$. The solid and dashed lines are $\eta\!=\!3$ and $\eta\!=\!2/5$, respectively, and the green shading is $8/3\!<\!\eta\!<\!3$, where $\eta\!=\!b_3^2/b_2b_4$. Energy profiles in three sub-regions are illustrated as follows: a single minimum for $\eta\!<\!8/3$ (bottom-right), a higher-energy second minimum for $8/3\!<\!\eta\!<\!3$ (top-left, green), and a lower-energy second minimum for $\eta\!>\!3$ (top-left, black). See text for discussion and Appendix~\ref{sec_activation_barrier:appendix_B} for details.}
\label{fig_activation_barrier:Fig1}
\end{figure}

We denote any local minimum of $\mathcal{E}(\bm z)$ by ${\bm \pi}$, to be termed  a ``cubic mode'', as it is a nonlinear mode defined through expansion coefficients up to cubic order. As $\mathcal{E}({\bm z})$ generally features multiple local minima, cubic modes obtained by minimizing it depend on the choice of initial trial directions. To exhaustively find all ${\bm \pi}$'s of a given minimum of $H$, the initial trial directions should be thoughtfully chosen, as discussed below and in Appendix~\ref{sec_activation_barrier:appendix_B}. We first discuss the shape parameter $\eta\=b_3^2/b_2b_4$ of cubic modes, and to this aim we scatter-plot in Fig.~\ref{fig_activation_barrier:Fig1} the distinct cubic modes of a single realization of the model with $N\=8192$ in the $b_3^2$--$\,b_2b_4$ plane, on a double-logarithmic scale. In this representation, lines of unit slope correspond to a fixed $\eta$ value.

The cubic modes form a well-defined band in the $b_3^2$--$b_2b_4$ plane, closely resembling the corresponding observation in finite-dimensional glasses (see, e.g., Fig.~11 of~\cite{Lerner2021-en}). Within this band, the parameter $\eta$ distinguishes three sub-regions according to the number and relative energies of the minima of $v(y;\eta)$ (see Fig.~\ref{fig_activation_barrier:Fig1}). Most modes reside in the $2/5\!<\!\eta\!<\!8/3$ sub-region, where the energy profile has only one minimum, at $y\!=\!0$, and hence features no second minimum for activated dynamics along this direction (the lower bound $\eta\!=\!2/5$, marked by the dashed line, emerges also in finite-dimensional glasses, but remains theoretically unexplained). A small fraction of the cubic modes resides in the $8/3\!<\!\eta\!<\!3$ sub-region, where the profile has a second minimum at a higher energy than the reference minimum at $y\!=\!0$. A nearly equal fraction resides in the $\eta\!>\!3$ sub-region, where the second minimum instead lies at a lower energy. We expect the sub-region with $8/3\!<\!\eta\!<\!3$ to be most relevant for activated dynamics, as further discussed below.

\begin{figure}[t!]
\centering
\includegraphics[width=0.97\columnwidth]{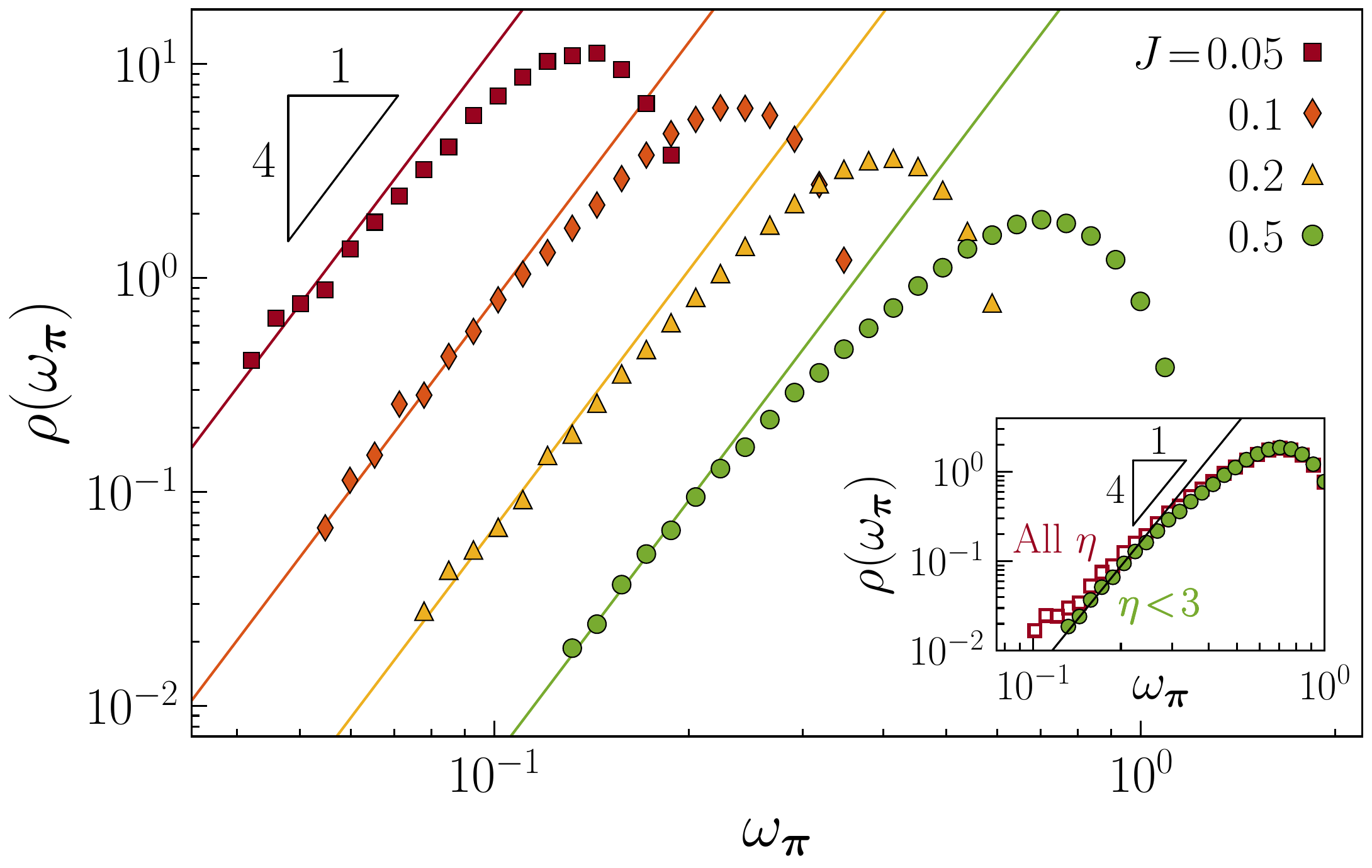}
\caption{The distributions $\rho(\omega_{\bm \pi})$ of cubic-mode frequency $\omega_{\bm \pi}\!=\!\sqrt{b_2(\bm \pi)}$ for $N\!=\!8192$ and various $J$ values (see legend), conditioned on $\eta\!<\!3$. (inset) $\rho(\omega_{\bm \pi})$ for $N\!=\!8192$ and $J\!=\!0.5$, with (green) or without (red) the condition $\eta\!<\!3$. The solid lines indicate $\rho(\omega_{\bm \pi})\!\sim\!\omega_{\bm \pi}^4$. See text for discussion, and Appendix~\ref{sec_activation_barrier:appendix_B} for details and related supporting results.}
\label{fig_acivation_barrier:Fig2}
\end{figure}

Having discussed the combination of expansion coefficients ${\cal E}$ and $\eta$, we consider next $b_2$ and $b_4$ individually. As the curvature $b_2(\bm \pi)$ defines the cubic-mode frequency through $\omega_{\bm \pi}\!\equiv\!\sqrt{b_2({\bm \pi})}$, we ask whether its distribution, $\rho(\omega_{\bm \pi})$, captures the salient properties of the glassy harmonic VDoS discussed above, i.e., a quartic low-frequency tail and a characteristic peak. Earlier work~\cite{Karpov1983-ri,Il-In1987-by,Buchenau1991-he,Gurarie2003-qi,Gurevich2003-oo,Gurevich2005-ru,Parshin2007-od,Bouchbinder2021-dh} predicted a quartic low-frequency tail while focusing on glassy minima with no deeper nearby minima. For cubic modes, an analogous condition amounts to $\eta\!<\!3$, which excludes modes with a deeper second minimum along the corresponding energy profile. Our numerical minimization of $H$, detailed in Appendix~\ref{sec_activation_barrier:appendix_F}, does not entirely exclude the existence of such deeper minima, such that a small fraction of the modes can feature $\eta\!>\!3$. Yet, we expect the condition $\eta\!<\!3$ to be physically realized in the presence of thermal fluctuations, which enable escape toward deeper minima. Moreover, modes with $\eta\!>\!3$ are expected to contribute only at very low frequencies, as is indeed demonstrated in the inset of Fig.~\ref{fig_acivation_barrier:Fig2}.

Consequently, we construct $\rho(\omega_{\bm \pi})$ in the main panel of Fig.~\ref{fig_acivation_barrier:Fig2} using only modes with $\eta\!<\!3$, and likewise use these modes hereafter. The resulting spectra exhibit the universal quartic tail, $\rho(\omega_{\bm \pi})\!\sim\!\omega_{\bm \pi}^4$, and a characteristic peak for all $J$ values considered. Moreover, the overall shape of $\rho(\omega_{\bm \pi})$ is $N$ independent (see Appendix~\ref{sec_activation_barrier:appendix_B}), indicating that our catalog of cubic modes is exhaustive. Finally, we find that $\langle\omega_{\bm \pi}\rangle\!\sim\!\omega_{\rm p}$ (where $\omega_{\rm p}$ is the intrinsic boson peak frequency defined above, see also~\cite{Suda2026-cl} and Appendix~\ref{sec_activation_barrier:appendix_B}) and that the ${\bm \pi}$'s are localized, i.e., feature a participation ratio of ${\cal O}(1/N)$ (see Appendix~\ref{sec_activation_barrier:appendix_B}). These results establish the correspondence between the anharmonic, cubic-mode spectra $\rho(\omega_{\bm \pi})$ and the harmonic VDoS $\mathcal{D}_{\!_{\rm G}\!}(\omega)$~\cite{Bouchbinder2021-dh,Rainone2021-am,Suda2026-cl,Moriel2024-wp}. As such, we hereafter do not distinguish between $\omega_{\bm \pi}$ and $\omega$ (and use the latter).

Finally, we consider $b_4$, which is shown to satisfy (see Appendix~\ref{sec_activation_barrier:appendix_A} for details)
\begin{equation}
    [b_4({\bm z})]^{-1} = N e({\bm z}) \ ,
\label{eq:b_4_PN}
\end{equation}
where $e({\bm z})\=(N\sum_i z_i^4)^{-1}$ is the participation ratio of a generalized direction ${\bm z}$ ($z_i$ is its $i^{\rm \tiny th}$ component). 
That is, $b_4({\bm z})$ is exactly the inverse of the participation number $N e({\bm z})$ of ${\bm z}$. The exact nature of Eq.~\eqref{eq:b_4_PN} stems from the quartic truncation of $H$ in Eq.~\eqref{eq:Hamiltonian}. Yet, a scaling relation that is consistent with Eq.~\eqref{eq:b_4_PN} has been numerically demonstrated in finite-dimensional glasses~\cite{Gartner2016-lc,Lerner2016-ra,Kapteijns2020-fu,Ji2022-os}, suggesting that quartic anharmonicity plays a dominant role in the localization of modes in realistic glasses, and that stochasticity in its amplitude (which we neglect in the model) is not essential~\cite{Parshin2007-od}. Applying then Eq.~\eqref{eq:b_4_PN} to cubic modes and recalling their localized nature (see above), we conclude that $b_4({\bm \pi})$ features a characteristic value independent of $\omega$, as explicitly demonstrated in Appendix~\ref{sec_activation_barrier:appendix_B}.\\

\section{T\lowercase{heory of activation barriers and plasticity onset}}
Next, we use the established properties of cubic modes in the model to derive theoretical predictions for the landscape's activation barriers $\Delta{U}$ and the external force $f_{\rm c}$ needed to trigger plasticity. As discussed above, cubic modes with $8/3\!<\!\eta\!<\!3$  single out landscape directions that are most relevant for activation, and hence we first aim at computing the statistics of activation barriers in these directions. These cubic modes also feature vibrational frequencies in the universal $\sim\!\omega^4$ tail. Furthermore, since $\eta\=b_3^2/b_2b_4$ features a characteristic $\omega$-independent value and the same applies to $b_4$, we expect
\begin{equation}
    |b_3|\sim\sqrt{b_2}\,.
\label{eq:b_3_b_2_relation}
\end{equation}

Since ${\cal E}\=b_2^3/b_3^2$ determines the scale of $\Delta{U}$, as explained above, Eq.~\eqref{eq:b_3_b_2_relation} immediately yields $\Delta{U}\!\sim\!b_2^2$. We expect this scaling relation to characterize the energy landscape, as further discussed below. Yet, we first focus on the {\it dynamical} questions of which landscape directions are spontaneously selected by the system/model under external forces and whether we can predict the landscape's properties in these directions. Specifically, applying the uniform external force $f(t)$ in Eq.~\eqref{eq:Hamiltonian}, starting from one of $H$'s local minima, will lead to a saddle-node bifurcation and to the associated fold catastrophe (plastic instability) at a certain critical force level $f_{\rm c}$. The latter marks the onset of plasticity/dissipation, bounding from above the elasticity domain of the model.

To predict $f_{\rm c}$ and its relation to the activation barrier $\Delta{U}_{\rm c}$ in the dynamically-selected direction, we assume that the uniform driving selects the activation of a plastic instability along one of the softest cubic modes. Under this assumption, we can readily relate $f_{\rm c}$ and $\Delta{U}_{\rm c}$ by analyzing the stationary points of $\delta{H}(s)\!-\!fs$, leading to $f_{\rm c}\!\sim\!b_2^2\,dy/ds\!\sim\!b_2|b_3|\!\sim\!b_2^{3/2}$, where again Eq.~\eqref{eq:b_3_b_2_relation} was used. Recalling that $b_2\=\omega^2$, and hence $\Delta{U}_{\rm c}\!\sim\!b_2^2\!\sim\!\omega^4$, we conclude that
\begin{equation}
   f_{\rm c} \sim (\Delta{U}_{\rm c})^{3/4}\sim \omega^3\,.
\label{eq:Delta_U_f_c_relations}
\end{equation}
Combined with ${\cal D}_{\!_{\rm G}\!}(\omega)\!\sim\!\omega^4$, we obtain the predictions (see Appendix~\ref{sec_activation_barrier:appendix_C} for details)
\begin{equation}
  p(\Delta{U}_{\rm c})\sim(\Delta{U}_{\rm c})^{1/4} \qquad\,\, \hbox{and}\qquad\,\, p(f_{\rm c})\sim f_{\rm c}^{2/3}\,,
\label{eq:Delta_U_f_c_stat}
\end{equation}
for the asymptotic statistics of $\Delta{U}_{\rm c}$ and $f_{\rm c}$. Finally, since $\Delta{U}_{\rm c}$ and $f_{\rm c}$ probe the softest landscape directions, we expect their typical values to inherit their finite $N$ dependence from the minimal frequencies of the vibrational spectrum. Consequently, applying standard extreme-value statistical considerations, we expect (see Appendix~\ref{sec_activation_barrier:appendix_C} for details)
\begin{equation}
    \langle \Delta{U}_{\rm c}\rangle \sim N^{-4/5} \qquad\,\, \hbox{and}\qquad\,\, \langle f_{\rm c}\rangle \sim N^{-3/5}\,.
\label{eq:Delta_U_f_c_min}
\end{equation}

The above predictions focused on landscape directions that are dynamically selected by the system under a uniform mechanical driving (in the absence of thermal fluctuations). As such, they reveal dynamically-relevant landscape properties. Yet, the prediction $\Delta{U}\!\sim\!b_2^2$ concerns a more general property of the landscape, independently of external forces. That is, we expect
\begin{equation}
  \langle\Delta{U}\rangle \sim \langle\omega\rangle_{_{\!\Delta{U}}}^4\ ,
\label{eq:Delta_U_w4}
\end{equation}
where the subscript in $\langle\,\cdot\,\rangle_{_{\!\Delta{U}}}$ emphasizes that the average is taken over cubic-mode directions in which a second minimum exists (and hence a barrier $\Delta{U}$ exists). This is to be distinguished from the mean frequency corresponding to $\rho(\omega_{\bm \pi})$ in Fig.~\ref{fig_acivation_barrier:Fig2}. Yet, Eq.~\eqref{eq:Delta_U_w4} concerns a finite portion of the barrier distribution, not limited to the very soft, dynamically-selected directions and their corresponding barriers $\Delta{U}_{\rm c}$ (see Appendix~\ref{sec_activation_barrier:appendix_C} for details).

While parts of Eqs.~\eqref{eq:Delta_U_f_c_relations}--\eqref{eq:Delta_U_w4} appeared in previous glass physics literature~\cite{Karpov1983-ri,Il-In1987-by,Buchenau1991-he,Ji2020-qv,Ji2022-os,Ji2019-uj}, with the most comprehensive discussion for computer glasses in~\cite{Kapteijns2020-fu}, these predictions have not been derived and presented in their entirety, as done above. Moreover, while these predictions are expected to equally apply to the mean-field model and to finite-dimensional glasses, they have been neither comprehensively tested nor verified so far.

To discuss the latter, it is useful to present Eqs.~\eqref{eq:Delta_U_f_c_stat} and~\eqref{eq:Delta_U_f_c_min} in terms of a single pseudogap exponent $\theta$, defined through $p(f_{\rm c})\!\sim\!f_{\rm c}^{\theta}$. In these terms, the remaining predictions are expressed as $p(\Delta{U}_{\rm c})\!\sim\!(\Delta{U}_{\rm c})^{(3\theta-1)/4}$, $\langle f_{\rm c}\rangle \!\sim\!N^{-1/(\theta+1)}$ and $\langle \Delta{U}_{\rm c}\rangle\!\sim\!N^{-4/[3(\theta+1)]}$, where we predict $\theta\=2/3$. The distance to the first plastic instability, quantified above through $f_{\rm c}$, has received considerable attention in the literature. Since it is typically the shear strain $\gamma$ that is externally controlled, the distribution $p(\gamma_{\rm c})\!\sim\!\gamma_{\rm c}^{\theta}$ is usually considered, which is a direct counterpart of $p(f_{\rm c})\!\sim\!f_{\rm c}^{\theta}$. The early numerical simulations of~\cite{Karmakar2010-ft} found $\theta\!\approx\!0.6$, in reasonable agreement with our prediction. The same work has also found $\langle \gamma_{\rm c}\rangle\!\sim\!N^{-0.62}$, which has been reinforced later in~\cite{Hentschel2015-hi,Lerner2018-ko}, in excellent agreement with the prediction $\langle f_{\rm c}\rangle \sim N^{-3/5}$. These numerical works preceded any theoretical predictions. Subsequent work, however, offered a variety of different predictions for $\theta$, along with different values reported in numerical simulations~\cite{Ruscher2020-gh,Shang2020-jq,Krishnan2022-ar}. Importantly, some theoretical and modeling works argued that $\theta$ is nonuniversal, e.g., that it depends on the glass preparation history~\cite{Ji2019-uj} and/or on space dimensionality~\cite{Lin2014-ov,Lin2016-jo}.

In view of the fundamental importance of activation barriers $\Delta{U}$ in glasses, various methods for probing them have been developed, including eigenvector-following methods such as the Activation-Relaxation Technique nouveau (ARTn)~\cite{Doye1997-cz,Malek2000-mc,Rodney2009-uw,Rodney2009-ff,Kallel2010-fe,Koziatek2013-wm,Swayamjyoti2014-pw,Fan2015-vi,Rodney2011-mo} and path optimization methods such as the Nudged Elastic Band (NEB) method, coupled to various techniques for finding potential energy minima~\cite{Doliwa2003-ow,Reinisch2005-me,Hamdan2014-lj,Damart2018-bp,Khomenko2020-gk,Khomenko2021-js,Mocanu2023-kv,Ciarella2023-zf,Pica-Ciamarra2024-xk,Ji2025-xk,Ghanem2026-ui}. These methods have been applied to various computer glass models and attempts to compute $p(\Delta{U})$ have been made, see the references above. Yet, no method has tested the predictions for $\Delta{U}_{\rm c}$ and $\langle\Delta{U}\rangle$ in Eqs.~\eqref{eq:Delta_U_f_c_stat}--\eqref{eq:Delta_U_w4}, and none has so far resolved the pseudogapped tail of $p(\Delta{U})$. Next, we will quantitatively test all the above-discussed predictions.

\section{N\lowercase{umerical confirmation of the predictions}}

\begin{figure}[t!]
\centering
\includegraphics[width=0.95\columnwidth]{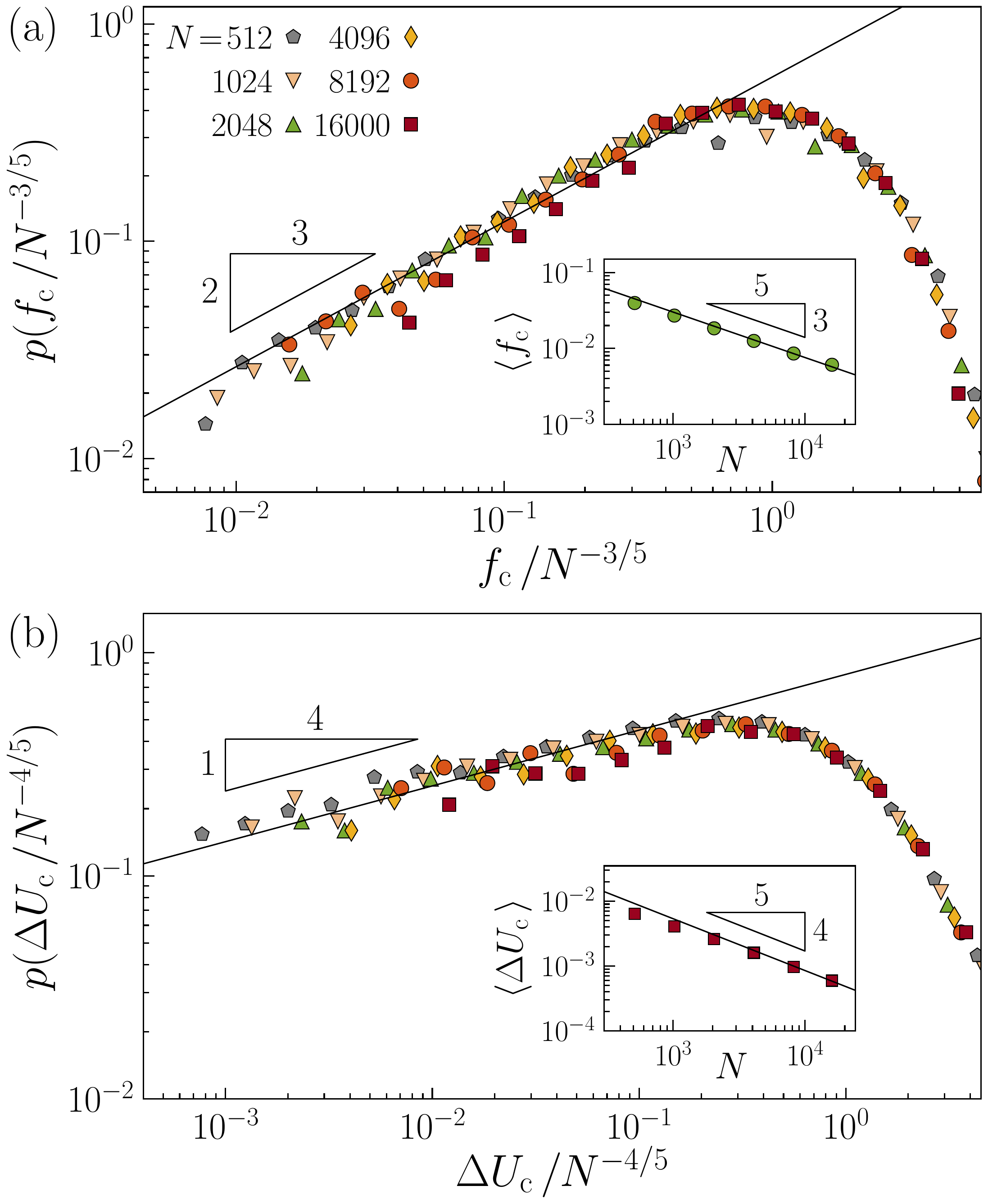}
\caption{
(a) The collapsed distribution $p(f_{\rm c}/N^{-3/5})$ of the distance to first instability for $J\!=\!0.5$ and various $N$ (see legend). (inset) The mean distance $\langle f_{\rm c}\rangle$ vs~$N$. (b) The collapsed distribution $p(\Delta{U}_{\rm c}/N^{-4/5})$ of the activation barrier for the same $J$ and $N$. (inset) The mean barrier height $\langle\Delta{U}_{\rm c}\rangle$ vs $N$. The solid lines are the power laws predicted in Eqs.~\eqref{eq:Delta_U_f_c_stat} and~\eqref{eq:Delta_U_f_c_min}; the triangles indicate their exponents. See text for discussion, and Appendices~\ref{sec_activation_barrier:appendix_D} and~\ref{sec_activation_barrier:appendix_E} for details.}
\label{fig_activation_barrier:Fig3}
\end{figure}

In numerically testing the predictions in Eqs.~\eqref{eq:Delta_U_f_c_stat}--\eqref{eq:Delta_U_w4}, one needs to distinguish between predictions associated with $f_{\rm c}$ and those that involve activation barriers $\Delta{U}$. Testing the former, in principle, is straightforward as it only requires applying a monotonically increasing external force $f(t)$ in Eq.~\eqref{eq:Hamiltonian}, starting at $t\=0$ from one of the minima of $H$, and to track the evolution of the system until a saddle-node bifurcation takes place. Pragmatically, this entails a careful selection of sufficiently small force increments, as done in established protocols~\cite{Ji2019-uj,Karmakar2010-ft,Hentschel2015-hi,Lerner2018-ko,Ruscher2020-gh,Shang2020-jq,Krishnan2022-ar}. Following these, we confirm in Fig.~\ref{fig_activation_barrier:Fig3}a the predictions $p(f_{\rm c})\sim f_{\rm c}^{2/3}$ and $\langle f_{\rm c}\rangle \sim N^{-3/5}$, see Appendix~\ref{sec_activation_barrier:appendix_E} for technical details.

Predictions involving activation barriers $\Delta{U}$ are significantly more challenging to test, for two main reasons. First, once a nearby minimum (relative to a reference minimum) is identified, it requires optimizing the landscape path connecting the two minima, to locate and resolve the saddle separating them. This task can be addressed using existing algorithms~\cite{Henkelman2000-hr,Olsen2004-px}, where we will specifically employ the NEB method for multi-dimensional landscape path optimization~\cite{Henkelman2000-hr}.

Second, the number of nearby minima is enormous (i.e., exponential in $N$) and most of them are irrelevant for activated dynamics. One therefore needs to identify a small subset of generalized landscape directions at a reference minimum and their associated barriers that are most relevant for crossing under thermomechanical driving. The cubic modes that were extensively discussed above --- most notably their softness, localization, and asymmetry --- constitute excellent candidates for such a physically-relevant subset of landscape directions.

In this respect, a clear advantage of the present model is its physically-motivated quartic truncation of $H$, which excludes higher-order anharmonicity present in finite-dimensional glasses (see discussion around Eq.~\eqref{eq:b_4_PN}). Moreover, since the model is exclusively formulated in terms of nonphononic (configurational) degrees of freedom, it bypasses the intrinsic difficulty encountered in finite-dimensional glasses, where nonphononic and phononic modes are hybridized/mixed in space~\cite{Lerner2023-dp}.

As noted in relation to $\rho(\omega_{\bm \pi})$ in Fig.~\ref{fig_acivation_barrier:Fig2} and as explained in detail in Appendix~\ref{sec_activation_barrier:appendix_B}, we developed a method to exhaustively find all distinct cubic modes $\bm \pi$ at a reference minimum of $H$. Having these at hand, we can find for each cubic mode with $8/3\!<\!\eta\!<\!3$ a nearby minimum of $H$ as described in Appendix~\ref{sec_activation_barrier:appendix_D}, and then employ the NEB method to locate the activation barrier separating the two minima of $H$. Applying this procedure to the softest cubic mode yields
$\Delta{U}_{\rm c}$, which is presented in Fig.~\ref{fig_activation_barrier:Fig3}b, fully confirming the predictions $p(\Delta{U}_{\rm c})\!\sim\!(\Delta{U}_{\rm c})^{1/4}$ and $\langle\Delta{U}_{\rm c}\rangle\!\sim\!N^{-4/5}$.
\begin{figure}[t!]
\centering
\includegraphics[width=0.95\columnwidth]{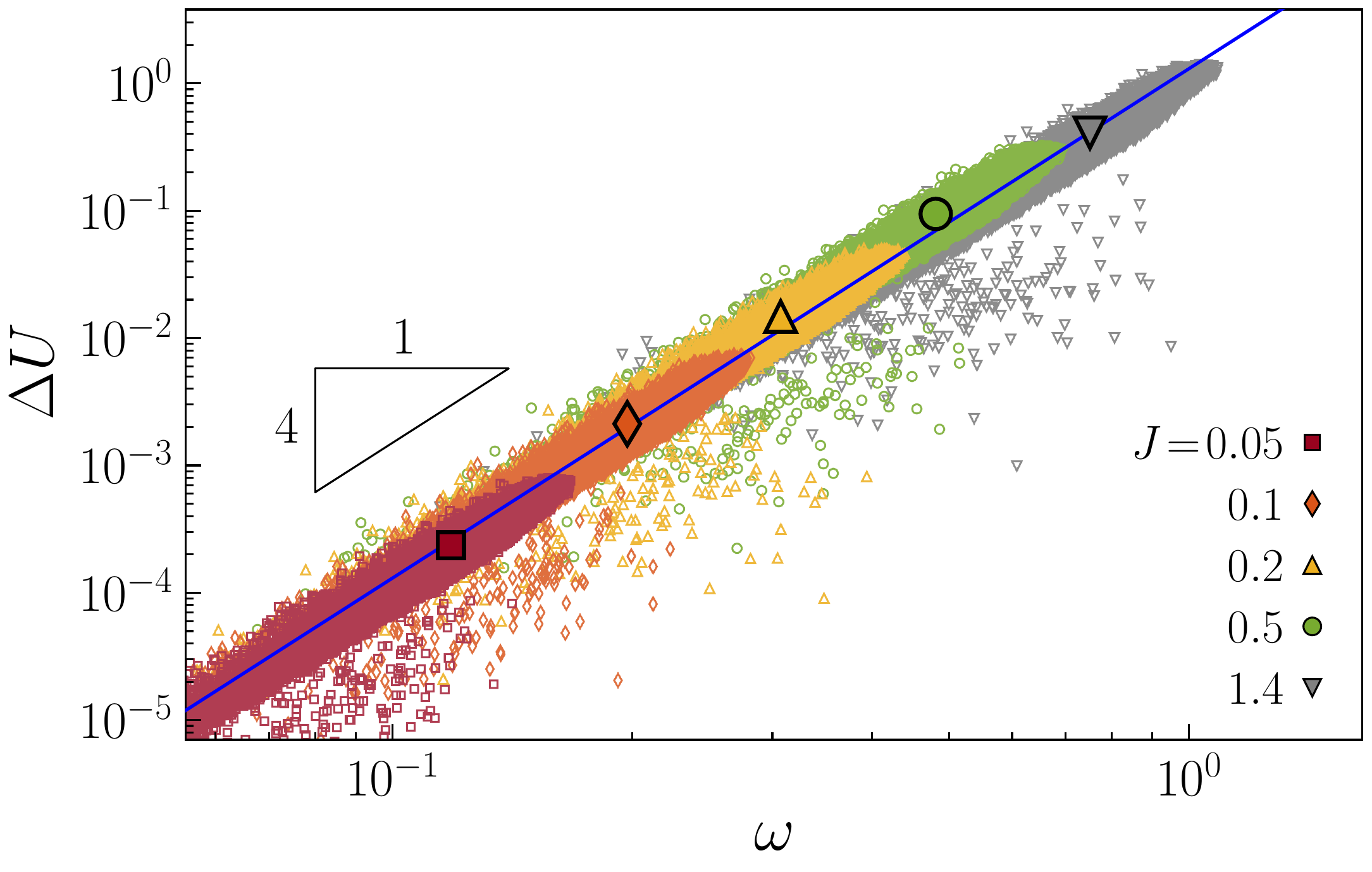}
\caption{The activation barrier $\Delta{U}$ vs~the corresponding cubic mode frequency $\omega$ for $N\!=\!8192$ and various $J$ values (see legend). For each $J$, the large black-outlined marker indicates the mean frequency and mean barrier height. The solid line is the power law $\Delta U\!\sim\!\omega^4$ predicted in Eq.~\eqref{eq:Delta_U_w4}. See text for discussion, and Appendix~\ref{sec_activation_barrier:appendix_D} for details.}
\label{fig_acivation_barrier:Fig4}
\end{figure}

Finally, we test Eq.~\eqref{eq:Delta_U_w4} by applying the same procedure to compute $\Delta{U}$ for all cubic modes $\bm \pi$ associated with a barrier, rather than only the softest. The results presented in Fig.~\ref{fig_acivation_barrier:Fig4} strongly support the theoretical prediction. We note that, in retrospect, earlier results for finite-dimensional computer glasses~\cite{Mizuno2020-uq} also provide indirect support for this prediction, see additional comments in~\footnote{Reference~\cite{Mizuno2020-uq} studied anharmonic glassy landscape properties along generalized directions corresponding to soft, quasi-localized eigenmodes, detecting and quantifying barriers along these directions. While these barriers, denoted by $\Delta\Phi_{\rm barrier}$ therein, are not associated with actual landscape saddles and thus differ from $\Delta{U}$ discussed in this work, the data presented in Figs.~3 and~5 of~\cite{Mizuno2020-uq} are scaling-wise consistent with Eqs.~\eqref{eq:Delta_U_f_c_relations} and~\eqref{eq:Delta_U_w4}, respectively.}.

\section{C\lowercase{oncluding remarks and outlook}}
In this work, we studied anharmonic properties of the energy landscape of a mean-field model of glasses, culminating in the analytic predictions in Eqs.~\eqref{eq:Delta_U_f_c_stat}--\eqref{eq:Delta_U_w4} for the statistics of activation barriers $\Delta{U}$ and the plasticity onset threshold $f_{\rm c}$, and their numerical confirmation. This progress is enabled by a nonlinear (cubic) modes framework, which is developed following its counterpart in finite-dimensional glasses. As the model predicts a broad range of glass properties and dynamical phenomena~\cite{Bouchbinder2021-dh,Rainone2021-am,Moriel2024-wp,Suda2025-kl,Suda2026-cl}, indicating the glassiness of its underlying landscape, the theoretical predictions in Eqs.~\eqref{eq:Delta_U_f_c_stat}--\eqref{eq:Delta_U_w4} are expected to be generic and hence valid also in finite-dimensional, realistic glasses.

The predictions for the asymptotic statistics of $f_{\rm c}$ and its extreme-value scaling are consistent with the earlier numerical results of~\cite{Karmakar2010-ft,Hentschel2015-hi,Lerner2018-ko} but at odds with other works, specifically with the claims that the pseudogap exponent $\theta$ may be non-universal. Future work should carefully test our prediction $\theta\=2/3$ in various computer glasses. Our predictions for the activation barriers $\Delta{U}$ have not been tested in finite-dimensional glasses so far, which is clearly a challenge for future work. Finally, accepting the generic nature of the predictions in Eqs.~\eqref{eq:Delta_U_f_c_stat}--\eqref{eq:Delta_U_w4}, an imminent next step is to understand their implications for thermo-mechanical activated dynamics, i.e., glassy relaxation at finite temperatures and glass deformation in response to external
forces larger than $f_{\rm c}$~\cite{Nagasawa2026-yr,Birks2026-wy,Ghanem2026-ui}.\\

\section*{A\lowercase{cknowledgements}}
E.B. acknowledges support from the Ben May Center for Chemical Theory and Computation and the Harold Perlman family. M.S. acknowledges support from the International Joint Graduate Program in Earth and Environmental Sciences, Tohoku University. The computations were carried out at Weizmann Institute of Science, using the Faculty of Chemistry’s high-performance computing facility CHEMFARM, which is supported in part by the Ben May Center for Chemical Theory and Computation.

\appendix
\setcounter{figure}{0}
\renewcommand{\thefigure}{A\arabic{figure}}
\section*{A\lowercase{ppendix}}

The goal of this Appendix is to provide technical details regarding findings presented in the main text and to offer some additional supporting results.

\section{T\lowercase{he coefficients $b_{2,3,4}$ in the main text's \uppercase{E}q.~(2) and derivation of \uppercase{E}q.~(4) therein}}
\label{sec_activation_barrier:appendix_A}

In Eq.~\eqref{eq:energy_expansion} of the main text, we introduce the expansion coefficients $b_{2,3,4}$, which are then used extensively. Here, we present their explicit expressions based on the one-dimensional energy profile $\delta H(s)\!\equiv\!H(\bm{x}_0+s\,\bm z)\!-\!H(\bm{x}_0)$, about a local minimum ${\bm x}_0$ of the Hamiltonian $H$ and along a generalized landscape direction ${\bm z}$. Since rescaling $\bm z$ merely reparametrizes $s$, we hereafter set $|\bm z|\!=\!1$. Consequently, we have
\begin{equation}
\begin{gathered}
    b_2(\bm z) \equiv \left.\frac{\pa^2 H}{\pa\bm x\pa\bm x}\right|_{\bm x = \bm x_0} : \bm z \bm z\,,\\
    b_3(\bm z) \equiv \left.\frac{\pa^3 H}{\pa\bm x\pa\bm x\pa\bm x}\right|_{\bm x = \bm x_0} :\!\cdot\, \bm z \bm z \bm z\,,\\
    b_4(\bm z) \equiv \left.\frac{\pa^4 H}{\pa\bm x\pa\bm x\pa\bm x\pa\bm x}\right|_{\bm x = \bm x_0} :: \bm z \bm z \bm z \bm z\,,
\end{gathered}
\label{eq:expansion_coeff_general}
\end{equation}
where $:,:\!\cdot,$ and $::$ denote double, triple, and quadruple tensor contractions, respectively. The general expressions in Eq.~\eqref{eq:expansion_coeff_general} are applied to our model using $H$ in Eq.~\eqref{eq:Hamiltonian} of the main text, from which the following partial derivatives are obtained
\begin{equation}
\begin{gathered}
    \frac{\pa^2 H}{\pa x_i\pa x_j} = \delta_{ij}(k_i + \tfrac{1}{2}x_i^2) + J_{ij}\,,\\
    \frac{\pa^3 H}{\pa x_i \pa x_j \pa x_k} =
    \begin{cases}
        x_i & (i\!=\!j\!=\!k)\\
        0 & ({\rm otherwise})
    \end{cases}\,,\\
    \frac{\pa^4 H}{\pa x_i \pa x_j \pa x_k \pa x_\ell} =
    \begin{cases}
        1 & (i\!=\!j\!=\!k\!=\!\ell)\\
        0 & ({\rm otherwise})
    \end{cases}\,,
\end{gathered}
\label{eq:partial_der_H}
\end{equation}
where $\delta_{ij}$ is the Kronecker delta and no summation convention is intended. Finally, the last expressions in Eqs.~\eqref{eq:expansion_coeff_general} and~\eqref{eq:partial_der_H} immediately yield
\begin{equation}
    b_4(\bm z) = \sum_i z_i^4\,,
\end{equation}
which is Eq.~\eqref{eq:b_4_PN} of the main text (see discussion therein and additional relevant results in Appendix~\ref{sec_activation_barrier:appendix_B}).

\section{A\lowercase{n exhaustive sampling of cubic modes and their properties}}
\label{sec_activation_barrier:appendix_B}

In Fig.~\ref{fig_acivation_barrier:Fig2} of the main text, we present the distributions $\rho(\omega_{\bm \pi})$ of cubic-mode frequencies $\omega_{\bm \pi}\=\sqrt{b_2({\bm \pi})}$ for a fixed value of $N$ and various $J$ values, and explain that they capture the main properties of the harmonic, non-Debye VDoS ${\cal D}_{\!_{\rm G}\!}(\omega)$. Moreover, we argue that our catalog of cubic modes is exhaustive, that $\langle\omega_{\bm \pi}\rangle\!\sim\!\omega_{\rm p}$ (where $\omega_{\rm p}$ is the local maximum of ${\cal D}_{\!_{\rm G}\!}(\omega)$), and that the ${\bm \pi}$'s are localized. Here, we quantitatively support these claims.

As explained in the main text, we obtain a cubic mode ${\bm \pi}$ by minimizing $\mathcal{E}(\bm z)$ of Eq.~\eqref{eq:cost_function} of the main text using standard numerical optimization techniques (described below) starting from an initial trial direction ${\bm z}$. Since $\mathcal{E}(\bm z)$ generally has multiple local minima, the cubic modes obtained depend on the choice of initial trial directions. For instance, previous studies used low-frequency eigenmodes~\cite{Kapteijns2020-fu} and local dipole responses~\cite{Richard2023-bo}, with the latter providing a systematic way to sample cubic modes throughout the glass. In this work, motivated by the recently identified correspondence between dipole response in finite-dimensional glasses and monopole responses in the present mean-field model~\cite{Suda2026-cl}, we adopt the latter strategy, and use the monopole responses as initial trial directions.

Specifically, consider a local force monopole $\bm{f}_i$ that is an $N$-dimensional force vector with a unit entry at the $i^{\rm \tiny th}$ oscillator and zeros elsewhere. Its linear displacement response, the so-called monopole response, reads ${\bm z}_i\!\equiv\!\bm{\mathcal{M}}^{-1}\cdot\bm{f}_i$, where $\bm{\mathcal{M}}\!=\!\pa^2H/\pa\bm{x}\pa\bm{x}$ is the Hessian evaluated at a reference minimum of $H$.
The properties of these responses and their correspondence with the finite-dimensional dipolar counterparts are discussed in~\cite{Suda2026-cl}. We then minimize $\mathcal{E}({\bm z})$ of Eq.~\eqref{eq:cost_function} of the main text starting from each of the $N$ monopole responses per minimum of $H$, which maps every ${\bm z}_i$ to a cubic mode ${\bm \pi}_i$. Different monopole responses can and indeed do converge to the same mode; retaining only distinct outcomes yields a catalog $\{\bm{\pi}_j\}_{j=1-\mathcal{N}}$, where $\mathcal{N}\!<\!N$ is the number of distinct modes found. The detailed properties of this mapping and the relations between the monopole responses and the resulting cubic modes will be extensively discussed elsewhere. Here, we focus next on assessing the statistical completeness of the resulting catalog of cubic modes by examining its dependence on the system size $N$.

\begin{figure}[t!]
\centering
\includegraphics[width=0.95\columnwidth]{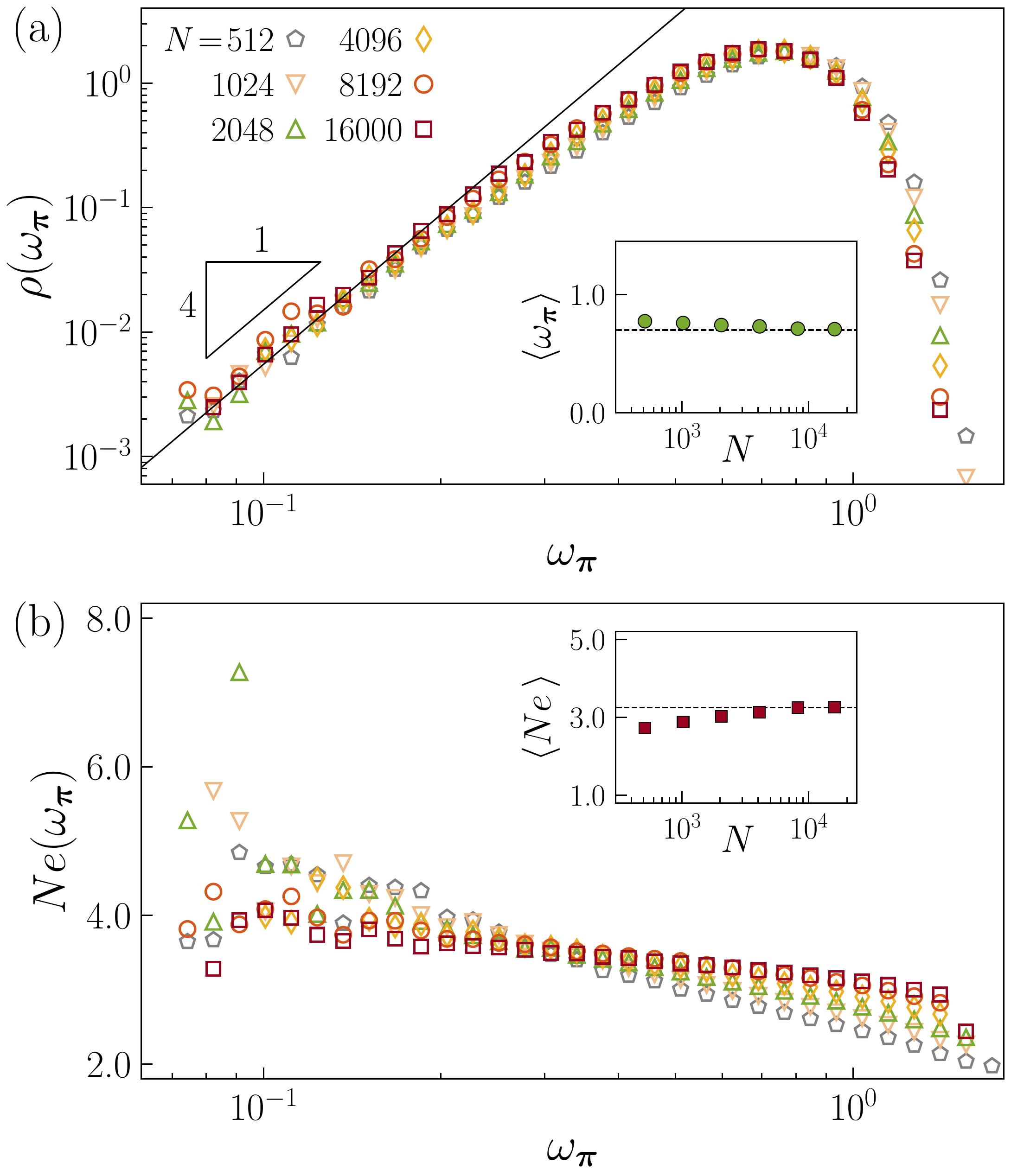}
\caption{(a) The distribution $\rho(\omega_{\bm \pi})$ of cubic mode frequency $\omega_{\bm \pi}\!=\!\sqrt{b_2(\bm \pi)}$ for $J\!=\!0.5$ and various $N$ (see legend). The solid line indicates a quartic low-frequency tail $\rho(\omega_{\bm \pi})\!\sim\!\omega_{\bm \pi}^4$. (inset) The mean cubic mode frequency $\langle \omega_{\bm \pi}\rangle$ vs $N$. The horizontal dashed line indicates an $N$-independent value $\langle\omega_{\bm \pi}\rangle\!\approx\!0.7$. (b) The frequency-averaged participation number of cubic modes, $Ne(\omega_{\bm \pi})$, for the same $J$ and $N$. (inset) the mean participation number $\langle Ne\rangle$ of cubic modes vs $N$. The horizontal dashed line indicates an $N$-independent value $\langle Ne\rangle\!\approx\!3.3$. The presented results are conditioned on $\eta\!<\!3$. See text for discussion.}
\label{fig_acivation_barrier:FigS1}
\end{figure}

\begin{figure}[t!]
\centering
\includegraphics[width=0.95\columnwidth]{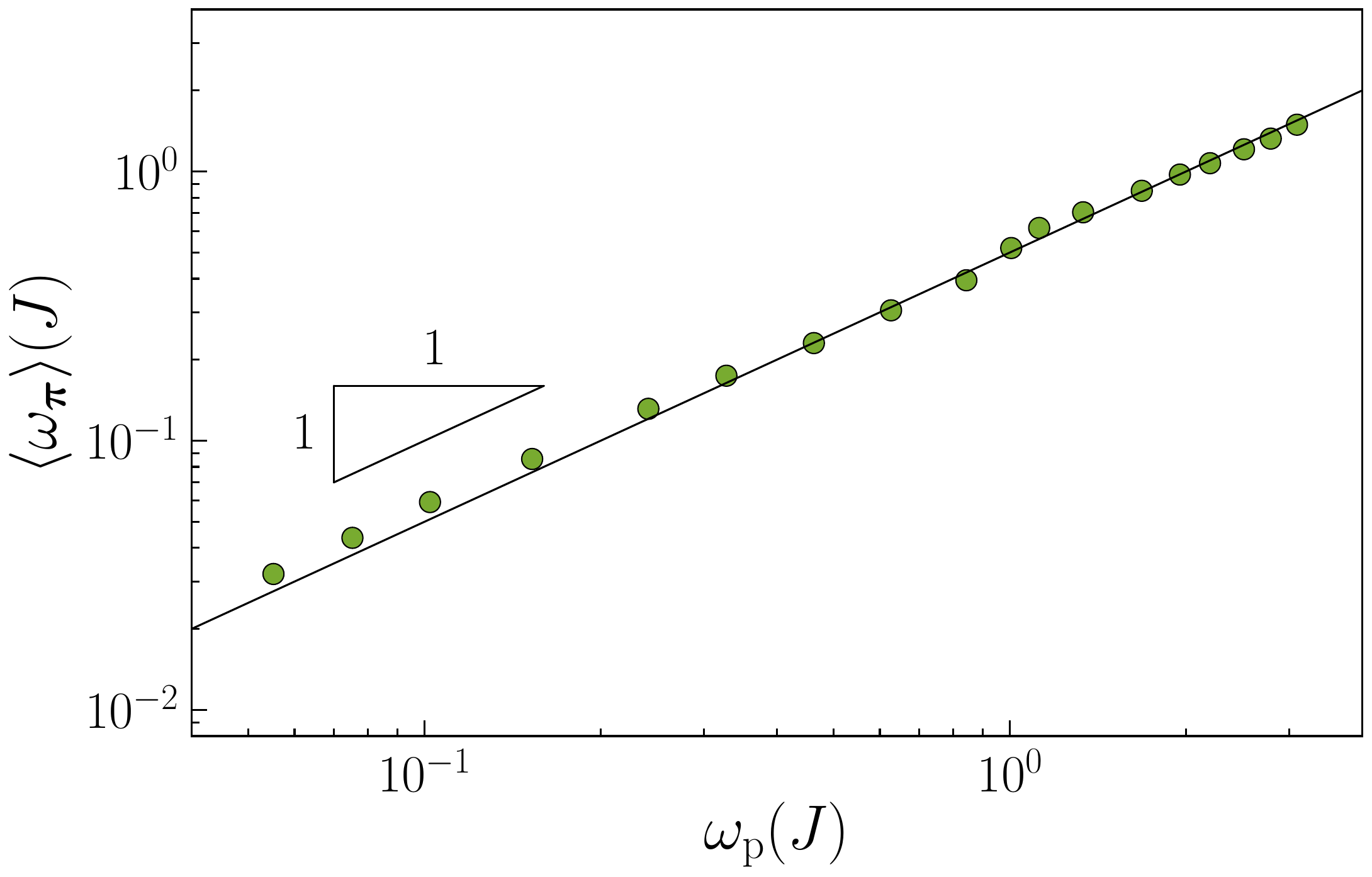}
\caption{The mean cubic mode frequency $\langle\omega_{\bm \pi}\rangle$ vs the intrinsic boson peak frequency $\omega_{\rm p}$~\cite{Suda2026-cl} for $N\!=\!8192$ and $0.01\!\leq\!J\!\leq\!2$. The solid line corresponds to $\langle\omega_{\bm \pi}\rangle\!\sim\!\omega_{\rm p}$. See text for discussion.}
\label{fig_acivation_barrier:FigS2}
\end{figure}

In Fig.~\ref{fig_acivation_barrier:FigS1}, we present the cubic mode frequency distributions $\rho(\omega_{\bm \pi})$ in panel (a) and the frequency-averaged participation number $Ne(\omega_{\bm \pi})$ in panel (b), both of which are conditioned on $\eta\!<\!3$, as motivated in the main text. Both profiles show no systematic dependence on $N$, as do their corresponding means $\langle\omega_{\bm \pi}\rangle$ and $\langle Ne\rangle$ shown in the insets. These results indicate the statistical completeness of the sampled cubic modes. Furthermore, the $N$-independence of $Ne$ in Fig.~\ref{fig_acivation_barrier:FigS1}b indicates that the cubic modes are localized. In fact, the weak frequency dependence of $Ne(\omega_{\bm \pi})$ therein, together with $b_2\!=\!\omega_{\bm \pi}^2$ and $[b_4(\bm \pi)]^{-1}\!=\!Ne(\bm \pi)$ (Eq.~(4) of the main text, also discussed above), indicates that $b_4(\bm \pi)$ does not systematically scale with $b_2(\bm \pi)$, as noted in the main text. Finally, we compare in Fig.~\ref{fig_acivation_barrier:FigS2} the mean cubic mode frequency $\langle\omega_{\bm \pi}\rangle$ and the intrinsic boson peak frequency $\omega_{\rm p}$ reported in~\cite{Suda2026-cl} across a range of disorder strengths $J$. It is observed that the two frequencies scale with one another. Taken together, the results presented in Figs.~\ref{fig_acivation_barrier:FigS1}--\ref{fig_acivation_barrier:FigS2} establish the localization and scaling properties of cubic modes discussed in the main text.

From a technical point of view, the minimization of $\mathcal{E}({\bm z})$, described above and used throughout the work, employs the FIRE algorithm~\cite{Bitzek2006-ht}, which was found to be stable and efficient, while maintaining $|\bm z|\!=\!1$. Since $\mathcal{E}({\bm z})$ is invariant under rescaling of ${\bm z}$ (i.e., as stressed, the latter is a generalized direction in the landscape), this normalization fixes the otherwise arbitrary magnitude of the trial direction. We drive the fictitious dynamics with a rescaled force $\bm{F}\!=\!-(b_3^2/6b_2^2)\pa\mathcal{E}/\pa\bm{z}$ and unit fictitious mass. The former rescaling removes a mode-dependent prefactor from the force magnitude, while preserving the descent direction. At each iteration, we also project the force and velocity onto the tangent space of a unit sphere, removing their components parallel to ${\bm z}$. We adopt the dimensionless FIRE parameters of~\cite{Bitzek2006-ht}, with an initial timestep $\Delta{t}_0\!=\!0.1(k_0+2J)^{-1/2}$ and a maximum timestep $\Delta{t}_{\rm max}\!=\!10\Delta{t}_0$. Here, $k_0+2J$ combines the maximum bare onsite stiffness $k_0$, which is set to $k_0\=1$ in this work, with the large-$N$ (spectral) contribution of the disordered interactions, providing a reference inertial timescale $(k_0+2J)^{-1/2}$. The prefactor $0.1$ sets a small initial step relative to this reference timescale, and FIRE subsequently adapts the timestep during minimization.

\section{T\lowercase{he main theoretical predictions in \uppercase{E}qs.~(5)--(9) of the main text}}
\label{sec_activation_barrier:appendix_C}

Here, we provide additional details regarding the derivation of the main theoretical predictions in Eqs.~\eqref{eq:b_3_b_2_relation}--\eqref{eq:Delta_U_w4} of the main text. As described therein, we start with Eq.~\eqref{eq:energy_expansion} and first separate the energy scale from the shape of the directional energy profile by introducing a dimensionless displacement $y\!\equiv\!s/(b_2/|b_3|)$. Choosing the orientation of $\bm z$ such that $b_3\!<\!0$, the profile reads
\begin{equation}
    \delta H(y) = \mathcal{E}\,v(y;\eta)\,,\qquad\quad \mathcal{E}=\!{b_2^3}/{b_3^2}\,,
\end{equation}
where the dimensionless profile $v(y;\eta)$ is given by
\begin{equation}
    v(y;\eta)\!\equiv\! y^2/2-y^3/3!+\eta^{-1}y^4/4!\,,\qquad\eta\!=\! {b_3^2}/{b_2b_4}\,.
\end{equation}
The shape parameter $\eta$ determines whether the profile contains a second minimum and whether its energy exceeds that of the reference minimum at $y\!=\!0$. That is, the stationary condition $v'(y;\eta)\!=\!0$ has two distinct nonzero solutions for $\eta\!>\!8/3$, at $y_\pm(\eta)\!=\! (3\eta\!\pm\!\sqrt{9\eta^2\!-\!24\eta})/2$. The smaller solution $y_{-}$ corresponds to the barrier, and the larger solution $y_{+}$ to the second minimum. Since $v(y_{\pm};\eta)\!=\!y_{\pm}^2(6\!-\!y_{\pm})/24$ and $v(y\=0;\eta)\=0$, the second minimum lies above the reference minimum for $y_{+}(\eta)\!<\!6$, i.e., for $8/3\!<\!\eta\!<\!3$. As discussed in the main text, we expect this interval to represent landscape directions relevant for the activated dynamics.

To identify directions with low energy barriers, we consider cubic modes $\bm \pi$. Since $b_4(\bm \pi)$ does not scale with $b_2(\bm \pi)$, as demonstrated above (and as have been demonstrated in finite-dimensional computer glasses~\cite{Gartner2016-lc,Kapteijns2020-fu}), for the physically-relevant interval $8/3\!<\!\eta\!<\!3$,  we obtain $|b_3|\!\sim\!\sqrt{b_2}$, which is Eq.~\eqref{eq:b_3_b_2_relation} of the main text. Using this scaling, the barrier height $\Delta U$ along a cubic mode follows directly from the energy profile,
\begin{equation}
    \Delta U \equiv (b_2^3/b_3^2)\,v(y_{-};\eta) \sim b_2^2\,,
\label{eq:deltaU_b2_SM}
\end{equation}
where we used the property that the dimensionless profile is bounded in the interval $9/8\!<\!v(y_{-};\eta)\!<\!4/3$ and that it does not scale with $b_2$. This relation concerns any cubic modes within the interval $8/3\!<\!\eta\!<\!3$, i.e., it corresponds to Eq.~\eqref{eq:Delta_U_w4} of the main text. We emphasize this relation is not limited to the very soft, dynamically-selected landscape directions which we discuss next.

When an external homogeneous driving $f(t)$ is imposed, the system dynamically selects a direction that leads to a saddle-node bifurcation. We approximate this direction by one of the softest cubic modes and denote the barrier along it as $\Delta{U}_{\rm c}$. We further approximate the force projected onto this direction by the applied external force $f(t)$, i.e., we neglect the fluctuations in the coupling between the homogeneous driving and individual cubic modes.

Within this approximation, the driven profile is $\delta H(s)\!-\!fs$, and the force $f_{\rm c}$ needed to eliminate the barrier is identified with the maximal restoring force between the reference minimum and the barrier at $y_{-}$. Evaluating this maximal force gives
\begin{equation}
    f_{\rm c} = \mathcal{E}\frac{dy}{ds}\max_{0\leq y\leq y_{\rm -}}{v'(y;\eta)} \sim b_2^{3/2}\,,
\label{eq:f_c_scaling_SM}
\end{equation}
where we used $\mathcal{E}\!\sim\!b_2^2$, $dy/ds\!=\!|b_3|/b_2\!\sim\!b_2^{-1/2}$, and that the maximum of $v'(y;\eta)$ remains finite and bounded away from zero throughout $8/3\!<\!\eta\!<\!3$ such that it introduces no additional scaling with $b_2$. Combining Eq.~\eqref{eq:deltaU_b2_SM} with Eq.~\eqref{eq:f_c_scaling_SM}, Eq.~\eqref{eq:Delta_U_f_c_relations} of the main text is obtained.

To derive the asymptotic statistical properties of these quantities, we connect the above relations to the glassy vibrational density of states (VDoS). To this aim, recall that the cubic-mode frequency distribution inherits the universal glassy low-frequency tail, $\rho(\omega_{\bm \pi})\!\sim\!\omega_{\bm \pi}^4$, as demonstrated in Fig.~\ref{fig_acivation_barrier:Fig2} of the main text. Then, since $\Delta{U}_{\rm c}\!\sim\!\omega_{\bm \pi}^4$, changing variables gives
\begin{equation}
    p(\Delta{U}_{\rm c})\sim \rho(\omega_{\bm \pi}) \left|\frac{d\omega_{\bm \pi}}{d\Delta{U}_{\rm c}}\right| \sim \omega_{\bm \pi}\sim (\Delta{U}_{\rm c})^{1/4}\,.
\end{equation}
and similarly, $f_{\rm c}\!\sim\!\omega_{\bm \pi}^3$ gives
\begin{equation}
    p(f_{\rm c})\sim \rho(\omega_{\bm \pi}) \left|\frac{d\omega_{\bm \pi}}{df_{\rm c}}\right| \sim \omega_{\bm \pi}^2 \sim f_{\rm c}^{2/3}\,,
\end{equation}
leading to Eq.~\eqref{eq:Delta_U_f_c_stat} of the main text. Unlike Eq.~\eqref{eq:deltaU_b2_SM}, the above relations describe the small-value tails of dynamically selected $\Delta{U}_{\rm c}$ and $f_{\rm c}$ distributions.

Finally, as external driving selects one of the softest landscape directions, we expect the typical values of $\Delta{U}_{\rm c}$ and $f_{\rm c}$ to inherit their finite $N$ dependence from the minimal frequencies of $\rho(\omega_{\bm \pi})$. Under this assumption, a standard extreme-value estimate determines the characteristic minimal frequency $\langle\omega_{\bm \pi,{\rm min}}\rangle$ through
\begin{equation}
    \int_0^{\langle\omega_{\bm \pi,{\rm min}}\rangle}\rho(\omega_{\bm \pi})d\omega_{\bm \pi} \sim N^{-1}.
\end{equation}
Combined with $\rho(\omega_{\bm \pi})\!\sim\!\omega_{\rm \pi}^4$, this gives $\langle\omega_{\bm \pi,{\rm min}}\rangle\!\sim\!N^{-1/5}$. Within this extreme-value estimate, the typical values of $\Delta{U}_{\rm c}$ and $f_{\rm c}$ are given by
\begin{equation}
\begin{split}
    \langle\Delta{U}_{\rm c}\rangle &\sim \langle\omega_{\bm \pi,{\rm min}}\rangle^4 \sim N^{-4/5}\,,\\
    \langle f_{\rm c}\rangle &\sim \langle\omega_{\bm \pi,{\rm min}}\rangle^3 \sim N^{-3/5}\,,
\end{split}
\end{equation}
which identify with Eq.~\eqref{eq:Delta_U_f_c_min} of the main text.

\section{S\lowercase{ampling energy barriers}}
\label{sec_activation_barrier:appendix_D}

In Figs.~\ref{fig_activation_barrier:Fig3}b and~\ref{fig_acivation_barrier:Fig4} of the main text, we present results regarding barriers in the model's energy landscape. Here, we provide additional details about the method employed in obtaining these results. As described in the main text, we use cubic modes to probe collective degrees of freedom relevant to activated dynamics. Our procedure consists of four steps: First, we select a cubic mode satisfying $8/3\!<\!\eta\!<\!3$. Second, we use its corresponding one-dimensional energy profile to locate another minimum along it. Third, while the latter is not a minimum of $H$, it typically resides in a basin of another minimum of $H$, which we identify through an unconstrained multi-dimensional minimization of $H$ (i.e., not constrained to the direction of the cubic mode) starting from the aforementioned, located one-dimensional energy minimum. Finally, we use the Climbing-Image Nudged Elastic Band (CI-NEB) method~\cite{Henkelman2000-hr} to locate a saddle connecting this minimum of $H$ to the reference minimum and determine the barrier height. We next describe these steps in more detail.

To select cubic modes, we first calculate the expansion coefficients $b_{2,3,4}$ along each monopole response $\bm z_i$. We minimize $\mathcal{E}(\bm z)$ starting from $\bm z_i$ only if $\eta(\bm{z}_i)\!>8/3$, so that its corresponding energy profile has a second minimum. This preliminary filter reduces the number of computationally expensive cubic-mode minimizations. We have verified that this filtering leaves the measured barrier statistics unchanged (additional properties of the mapping $\bm{z}_i\!\to\!\bm{\pi}_i$ will be discussed elsewhere). From the resulting cubic modes, we retain only those satisfying $8/3\!<\!\eta(\bm{\pi}_i)\!<\!3$, for which the second minimum lies above the reference minimum. To sample $\Delta{U}_{\rm c}$, we select the monopole response with the smallest associated stiffness $b_2(\bm{z}_i)$ and minimize $\mathcal{E}({\bm z})$ starting from this response, without applying the preliminary filter on $\eta(\bm{z}_i)$. We retain the resulting cubic mode only if it satisfies the same condition $8/3\!<\!\eta(\bm{\pi}_i)\!<\!3$.

For each retained cubic mode $\bm{\pi}_i$, we displace the reference configuration $\bm{x}_0$ to the second minimum of the corresponding one-dimensional energy profile. Orienting $\bm{\pi}_i$ such that $b_3(\bm{\pi}_i)\!<\!0$ and enforcing $|\bm{\pi}_i|\!=\!1$, as already noted above, we construct a trial configuration
\begin{equation}
    \bm{x}_{\rm trial} = \bm x_0 + s_{+}\bm{\pi}_i\,,
\end{equation}
where the scalar displacement $s_+$ is given by
\begin{equation}
    s_+ = \frac{b_2}{|b_3|} \frac{3\eta+\sqrt{9\eta^2-24\eta}}{2}\,.
\end{equation}
We subsequently minimize $H$ over all coordinates, without any directional or other constraint, starting from $\bm{x}_{\rm trial}$. The trial displacement $s_+$ is thus determined by the cubic mode's expansion coefficients alone, without introducing any adjustable parameters controlling the displacement amplitude. If the resulting minimum differs from $\bm{x}_0$ and from all previously retained minima, we add it to the catalog of nearby minima. We repeat this procedure for all retained cubic modes.

We optimize paths connecting $\bm{x}_0$ to each distinct nearby minimum using CI-NEB. We relax the band with the FIRE algorithm~\cite{Bitzek2006-ht}, using the same FIRE parameters as for the cubic-mode minimization described above. After convergence, we verify that the Hessian at the proposed saddle $\bm{x}_{\rm saddle}$ has exactly one negative eigenvalue and that minimizations initiated on opposite sides along its unstable eigenmode return to the two endpoint minima, one on each side. We retain only saddles that pass both checks and calculate the barrier height for each as $\Delta U\!=\!H(\bm{x}_{\rm saddle})\!-\!H(\bm{x}_0)$. Attempts that do not converge or fail saddle verification are discarded.

For CI-NEB, we use eight intermediate images per path, excluding the two fixed endpoints, and a uniform spring stiffness $k_{\rm NEB}\=0.3(k_0+2J)$. Here, $k_0+2J$ is the reference stiffness introduced above (and recall that we set $k_0\=1$ throughout), and the prefactor $0.3$ is selected empirically for rapid convergence. We have verified that increasing the number of images or varying the spring stiffness prefactor leaves the measured barrier statistics unchanged. To initialize each path, we use the cubic mode $\bm\pi_i$ that discovered the corresponding nearby landscape minimum. We place one of the central images at the saddle of $\bm{\pi}_i$'s energy profile, $\bm x_0\!+\!s_-\bm\pi_i$, where $s_-\!=\!(b_2/|b_3|)(3\eta\!-\!\sqrt{9\eta^2\!-\!24\eta})/2$. We find that this initialization accelerates convergence and increases the success rate of CI-NEB. For the endpoint-connectivity check, we minimize $H$ from $\bm x_{\rm saddle}\!\pm\!10^{-2}\bm\psi_-$, where $\bm\psi_-$ is the normalized unstable eigenmode.

\section{S\lowercase{ampling the first plastic instability}}
\label{sec_activation_barrier:appendix_E}

In Fig.~\ref{fig_activation_barrier:Fig3}a of the main text, we present results regarding the external force $f_{\rm c}$ required to reach the first plastic instability from an as-quenched configuration. Here, we describe how $f_{\rm c}$ is measured. Following the instability-detection algorithm developed in~\cite{Ji2019-uj}, we measure $f_{\rm c}$ by increasing the external force from $f(t)\!=\!0$ following an athermal quasi-static (AQS) loading protocol, where $H$ is minimized after each force increment $\Delta{f}$. To detect discontinuous oscillator motion associated with plastic instabilities, we monitor the largest oscillator displacement per unit force increment, $Q\!\equiv\!\max_{i}{|\Delta{x}_i|/\Delta{f}}$, where $\Delta{x}_i$ is the displacement of the $i^{\rm \tiny th}$ oscillator between consecutive energy-minimized configurations. At a fixed force on a smooth elastic branch, $\Delta x_i\!\approx\!(dx_i/df)\Delta f$ for sufficiently small $\Delta{f}$, and hence $Q$ approaches a finite value $\max_i|dx_i/df|$. For an increment crossing a plastic instability, the maximum displacement remains at a finite value even as $\Delta{f}$ is reduced, giving $Q\!\sim\!(\Delta{f})^{-1}$. We therefore track consecutive values of $Q$ to detect its sharp peaks as candidates for plastic instabilities.

Specifically, let $Q_0$, $Q_1$ and $Q_2$ denote three consecutive values of $Q$, with $Q_1$ corresponding to the increment from $f(t)$ to $f(t)\!+\!\Delta{f}$. We identify this increment as containing a candidate plastic instability if $Q_1\!>Q_0,Q_2$ and $Q_1\!>\!Q_{\rm c}$, where $Q_{\rm c}$ is an empirical threshold. Upon detecting a candidate instability, we restore the configuration at $f(t)$, reduce $\Delta{f}$ by a factor of ten, and repeat the detection procedure. Starting from $\Delta{f}_{\rm ini}$, we repeat this refinement until the target resolution $\Delta{f}_{\rm fin}$ is reached. At this final resolution, we additionally require $Q_1\!>\!100 Q_2$ as an operational criterion for distinguishing discontinuous plastic instabilities from large continuous responses.

For the results presented in Fig.~\ref{fig_activation_barrier:Fig3}a of the main text, we set $Q_{\rm c}\!=\!300$, $\Delta{f}_{\rm ini}\!=\!10^{-3}$, and $\Delta{f}_{\rm fin}\!=\!10^{-6}$. We verified that the chosen threshold avoids classifying large continuous non-affine responses as plastic instabilities, and that the detected instabilities are insensitive to further reductions in either $\Delta{f}_{\rm ini}$ or $\Delta{f}_{\rm fin}$. At the initial, as-quenched configuration, we also measure the shape parameter $\eta\!=\!b_3^2/b_2b_4$ of the softest cubic mode, obtained by minimizing $\mathcal{E}(\bm z)$ starting from the softest monopole response, as described above. This measurement allows us to restrict the $f_{\rm c}$ statistics a posteriori to realizations whose softest cubic mode satisfies $\eta\!<\!3$, as motivated in the main text.

\section{N\lowercase{umerical implementation of the mean-field model and additional details}}
\label{sec_activation_barrier:appendix_F}

The mean-field model we studied is formulated through the disordered Hamiltonian $H$ defined in Eq.~\eqref{eq:Hamiltonian} of the main text. For all the numerical results reported in this work, samples were prepared by performing energy minimization in the absence of an external force, $f(t)\!=\!0$, starting from random oscillator configurations where each $x_i$ is drawn from a uniform distribution over $[-\epsilon/2,\epsilon/2]$ with $\epsilon\!=\!10^{-2}$. We verified that the value of $\epsilon$ does not affect the results.

The number $M$ of independent quenched disorder realizations prepared for each figure is as follows. For Fig.~\ref{fig_activation_barrier:Fig1}, we prepared $M\!=\!1$ realization at $N\!=\!8192$. For Fig.~\ref{fig_acivation_barrier:Fig2}, we prepared $M\!=\!500$ realizations at $N\!=\!8192$ for each $J$ value. For Fig.~\ref{fig_activation_barrier:Fig3}a, we prepared $M\!=\!\{640,320,160,80,40,20\}\!\times\!10^3$ realizations for $N\!=\!\{512,1024,2048,4096,8192,16000\}$. For Fig.~\ref{fig_activation_barrier:Fig3}b, we prepared $M\!=\!\{1024,768,512,256,800,200\}\!\times\!10^3$ realizations for the same system sizes. For Fig.~\ref{fig_acivation_barrier:Fig4}, we prepared $M\!\approx\!2\!\times\!10^4$ realizations at $N\!=\!8192$ for each $J$ value. For Fig.~\ref{fig_acivation_barrier:FigS1}, we prepared $M\!=\!\{8000,4000,2000,1000,500,198\}$ realizations for $N\!=\!\{512,1024,2048,4096,8192,16000\}$. For Fig.~\ref{fig_acivation_barrier:FigS2}, at each $J$ value, we prepared $M\!\geq\!200$ realizations to calculate $\langle\omega_{\bm \pi}\rangle(J)$ and $M\!=\!2000$ realizations to calculate $\omega_{\rm p}(J)$, both at $N\!=\!8192$.


\begin{thebibliography}{94}%
\makeatletter
\providecommand \@ifxundefined [1]{%
 \@ifx{#1\undefined}
}%
\providecommand \@ifnum [1]{%
 \ifnum #1\expandafter \@firstoftwo
 \else \expandafter \@secondoftwo
 \fi
}%
\providecommand \@ifx [1]{%
 \ifx #1\expandafter \@firstoftwo
 \else \expandafter \@secondoftwo
 \fi
}%
\providecommand \natexlab [1]{#1}%
\providecommand \enquote  [1]{``#1''}%
\providecommand \bibnamefont  [1]{#1}%
\providecommand \bibfnamefont [1]{#1}%
\providecommand \citenamefont [1]{#1}%
\providecommand \href@noop [0]{\@secondoftwo}%
\providecommand \href [0]{\begingroup \@sanitize@url \@href}%
\providecommand \@href[1]{\@@startlink{#1}\@@href}%
\providecommand \@@href[1]{\endgroup#1\@@endlink}%
\providecommand \@sanitize@url [0]{\catcode `\\12\catcode `\$12\catcode
  `\&12\catcode `\#12\catcode `\^12\catcode `\_12\catcode `\%12\relax}%
\providecommand \@@startlink[1]{}%
\providecommand \@@endlink[0]{}%
\providecommand \url  [0]{\begingroup\@sanitize@url \@url }%
\providecommand \@url [1]{\endgroup\@href {#1}{\urlprefix }}%
\providecommand \urlprefix  [0]{URL }%
\providecommand \Eprint [0]{\href }%
\providecommand \doibase [0]{https://doi.org/}%
\providecommand \selectlanguage [0]{\@gobble}%
\providecommand \bibinfo  [0]{\@secondoftwo}%
\providecommand \bibfield  [0]{\@secondoftwo}%
\providecommand \translation [1]{[#1]}%
\providecommand \BibitemOpen [0]{}%
\providecommand \bibitemStop [0]{}%
\providecommand \bibitemNoStop [0]{.\EOS\space}%
\providecommand \EOS [0]{\spacefactor3000\relax}%
\providecommand \BibitemShut  [1]{\csname bibitem#1\endcsname}%
\let\auto@bib@innerbib\@empty
\bibitem [{\citenamefont {Kühn}\ and\ \citenamefont
  {Horstmann}(1997)}]{Kuhn1997-bl}%
  \BibitemOpen
  \bibfield  {author} {\bibinfo {author} {\bibfnamefont {R.}~\bibnamefont
  {Kühn}}\ and\ \bibinfo {author} {\bibfnamefont {U.}~\bibnamefont
  {Horstmann}},\ }\bibfield  {title} {\bibinfo {title} {Random matrix approach
  to glassy physics: Low temperatures and beyond},\ }\href
  {https://doi.org/10.1103/physrevlett.78.4067} {\bibfield  {journal} {\bibinfo
   {journal} {Phys. Rev. Lett.}\ }\textbf {\bibinfo {volume} {78}},\ \bibinfo
  {pages} {4067} (\bibinfo {year} {1997})}\BibitemShut {NoStop}%
\bibitem [{\citenamefont {Gurarie}\ and\ \citenamefont
  {Chalker}(2003)}]{Gurarie2003-qi}%
  \BibitemOpen
  \bibfield  {author} {\bibinfo {author} {\bibfnamefont {V.}~\bibnamefont
  {Gurarie}}\ and\ \bibinfo {author} {\bibfnamefont {J.~T.}\ \bibnamefont
  {Chalker}},\ }\bibfield  {title} {\bibinfo {title} {Bosonic excitations in
  random media},\ }\href {https://doi.org/10.1103/PhysRevB.68.134207}
  {\bibfield  {journal} {\bibinfo  {journal} {Phys. Rev. B}\ }\textbf {\bibinfo
  {volume} {68}},\ \bibinfo {pages} {134207} (\bibinfo {year}
  {2003})}\BibitemShut {NoStop}%
\bibitem [{\citenamefont {Gurevich}\ \emph {et~al.}(2003)\citenamefont
  {Gurevich}, \citenamefont {Parshin},\ and\ \citenamefont
  {Schober}}]{Gurevich2003-oo}%
  \BibitemOpen
  \bibfield  {author} {\bibinfo {author} {\bibfnamefont {V.~L.}\ \bibnamefont
  {Gurevich}}, \bibinfo {author} {\bibfnamefont {D.~A.}\ \bibnamefont
  {Parshin}},\ and\ \bibinfo {author} {\bibfnamefont {H.~R.}\ \bibnamefont
  {Schober}},\ }\bibfield  {title} {\bibinfo {title} {Anharmonicity,
  vibrational instability, and the boson peak in glasses},\ }\href
  {https://doi.org/10.1103/PhysRevB.67.094203} {\bibfield  {journal} {\bibinfo
  {journal} {Phys. Rev. B}\ }\textbf {\bibinfo {volume} {67}},\ \bibinfo
  {pages} {094203} (\bibinfo {year} {2003})}\BibitemShut {NoStop}%
\bibitem [{\citenamefont {Gurevich}\ \emph {et~al.}(2005)\citenamefont
  {Gurevich}, \citenamefont {Parshin},\ and\ \citenamefont
  {Schober}}]{Gurevich2005-ru}%
  \BibitemOpen
  \bibfield  {author} {\bibinfo {author} {\bibfnamefont {V.~L.}\ \bibnamefont
  {Gurevich}}, \bibinfo {author} {\bibfnamefont {D.~A.}\ \bibnamefont
  {Parshin}},\ and\ \bibinfo {author} {\bibfnamefont {H.~R.}\ \bibnamefont
  {Schober}},\ }\bibfield  {title} {\bibinfo {title} {Pressure dependence of
  the boson peak in glasses},\ }\href
  {https://doi.org/10.1103/physrevb.71.014209} {\bibfield  {journal} {\bibinfo
  {journal} {Phys. Rev. B}\ }\textbf {\bibinfo
  {volume} {71}},\ \bibinfo {pages} {014209} (\bibinfo {year}
  {2005})}\BibitemShut {NoStop}%
\bibitem [{\citenamefont {Parshin}\ \emph {et~al.}(2007)\citenamefont
  {Parshin}, \citenamefont {Schober},\ and\ \citenamefont
  {Gurevich}}]{Parshin2007-od}%
  \BibitemOpen
  \bibfield  {author} {\bibinfo {author} {\bibfnamefont {D.~A.}\ \bibnamefont
  {Parshin}}, \bibinfo {author} {\bibfnamefont {H.~R.}\ \bibnamefont
  {Schober}},\ and\ \bibinfo {author} {\bibfnamefont {V.~L.}\ \bibnamefont
  {Gurevich}},\ }\bibfield  {title} {\bibinfo {title} {Vibrational instability,
  two-level systems, and the boson peak in glasses},\ }\href
  {https://doi.org/10.1103/PhysRevB.76.064206} {\bibfield  {journal} {\bibinfo
  {journal} {Phys. Rev. B}\ }\textbf {\bibinfo {volume} {76}},\ \bibinfo
  {pages} {064206} (\bibinfo {year} {2007})}\BibitemShut {NoStop}%
\bibitem [{\citenamefont {Ritort}\ and\ \citenamefont
  {Sollich}(2003)}]{Ritort2003-yu}%
  \BibitemOpen
  \bibfield  {author} {\bibinfo {author} {\bibfnamefont {F.}~\bibnamefont
  {Ritort}}\ and\ \bibinfo {author} {\bibfnamefont {P.}~\bibnamefont
  {Sollich}},\ }\bibfield  {title} {\bibinfo {title} {Glassy dynamics of
  kinetically constrained models},\ }\href
  {https://doi.org/10.1080/0001873031000093582} {\bibfield  {journal} {\bibinfo
   {journal} {Adv. Phys.}\ }\textbf {\bibinfo {volume} {52}},\ \bibinfo {pages}
  {219} (\bibinfo {year} {2003})}\BibitemShut {NoStop}%
\bibitem [{\citenamefont {Cavagna}(2009)}]{Cavagna2009-wq}%
  \BibitemOpen
  \bibfield  {author} {\bibinfo {author} {\bibfnamefont {A.}~\bibnamefont
  {Cavagna}},\ }\bibfield  {title} {\bibinfo {title} {Supercooled liquids for
  pedestrians},\ }\href {https://doi.org/10.1016/j.physrep.2009.03.003}
  {\bibfield  {journal} {\bibinfo  {journal} {Phys. Rep.}\ }\textbf {\bibinfo
  {volume} {476}},\ \bibinfo {pages} {51} (\bibinfo {year} {2009})}\BibitemShut
  {NoStop}%
\bibitem [{\citenamefont {DeGiuli}\ \emph {et~al.}(2014)\citenamefont
  {DeGiuli}, \citenamefont {Laversanne-Finot}, \citenamefont {Düring},
  \citenamefont {Lerner},\ and\ \citenamefont {Wyart}}]{DeGiuli2014-oi}%
  \BibitemOpen
  \bibfield  {author} {\bibinfo {author} {\bibfnamefont {E.}~\bibnamefont
  {DeGiuli}}, \bibinfo {author} {\bibfnamefont {A.}~\bibnamefont
  {Laversanne-Finot}}, \bibinfo {author} {\bibfnamefont {G.}~\bibnamefont
  {Düring}}, \bibinfo {author} {\bibfnamefont {E.}~\bibnamefont {Lerner}},\
  and\ \bibinfo {author} {\bibfnamefont {M.}~\bibnamefont {Wyart}},\ }\bibfield
   {title} {\bibinfo {title} {Effects of coordination and pressure on sound
  attenuation, boson peak and elasticity in amorphous solids},\ }\href
  {https://doi.org/10.1039/c4sm00561a} {\bibfield  {journal} {\bibinfo
  {journal} {Soft Matter}\ }\textbf {\bibinfo {volume} {10}},\ \bibinfo {pages}
  {5628} (\bibinfo {year} {2014})}\BibitemShut {NoStop}%
\bibitem [{\citenamefont {Franz}\ \emph {et~al.}(2015)\citenamefont {Franz},
  \citenamefont {Parisi}, \citenamefont {Urbani},\ and\ \citenamefont
  {Zamponi}}]{Franz2015-cc}%
  \BibitemOpen
  \bibfield  {author} {\bibinfo {author} {\bibfnamefont {S.}~\bibnamefont
  {Franz}}, \bibinfo {author} {\bibfnamefont {G.}~\bibnamefont {Parisi}},
  \bibinfo {author} {\bibfnamefont {P.}~\bibnamefont {Urbani}},\ and\ \bibinfo
  {author} {\bibfnamefont {F.}~\bibnamefont {Zamponi}},\ }\bibfield  {title}
  {\bibinfo {title} {Universal spectrum of normal modes in low-temperature
  glasses},\ }\href {https://doi.org/10.1073/pnas.1511134112} {\bibfield
  {journal} {\bibinfo  {journal} {Proc. Natl. Acad. Sci. U. S. A.}\ }\textbf
  {\bibinfo {volume} {112}},\ \bibinfo {pages} {14539} (\bibinfo {year}
  {2015})}\BibitemShut {NoStop}%
\bibitem [{\citenamefont {Ros}\ \emph {et~al.}(2019)\citenamefont {Ros},
  \citenamefont {Ben~Arous}, \citenamefont {Biroli},\ and\ \citenamefont
  {Cammarota}}]{Ros2019-nl}%
  \BibitemOpen
  \bibfield  {author} {\bibinfo {author} {\bibfnamefont {V.}~\bibnamefont
  {Ros}}, \bibinfo {author} {\bibfnamefont {G.}~\bibnamefont {Ben~Arous}},
  \bibinfo {author} {\bibfnamefont {G.}~\bibnamefont {Biroli}},\ and\ \bibinfo
  {author} {\bibfnamefont {C.}~\bibnamefont {Cammarota}},\ }\bibfield  {title}
  {\bibinfo {title} {Complex energy landscapes in spiked-tensor and simple
  glassy models: Ruggedness, arrangements of local minima, and phase
  transitions},\ }\href {https://doi.org/10.1103/PhysRevX.9.011003} {\bibfield
  {journal} {\bibinfo  {journal} {Phys. Rev. X}\ }\textbf {\bibinfo {volume}
  {9}},\ \bibinfo {pages} {011003} (\bibinfo {year} {2019})}\BibitemShut
  {NoStop}%
\bibitem [{\citenamefont {Folena}\ \emph {et~al.}(2020)\citenamefont {Folena},
  \citenamefont {Franz},\ and\ \citenamefont
  {Ricci-Tersenghi}}]{Folena2020-ud}%
  \BibitemOpen
  \bibfield  {author} {\bibinfo {author} {\bibfnamefont {G.}~\bibnamefont
  {Folena}}, \bibinfo {author} {\bibfnamefont {S.}~\bibnamefont {Franz}},\ and\
  \bibinfo {author} {\bibfnamefont {F.}~\bibnamefont {Ricci-Tersenghi}},\
  }\bibfield  {title} {\bibinfo {title} {Rethinking mean-field glassy dynamics
  and its relation with the energy landscape: The surprising case of the
  spherical mixed p-spin model},\ }\href
  {https://doi.org/10.1103/physrevx.10.031045} {\bibfield  {journal} {\bibinfo
  {journal} {Phys. Rev. X}\ }\textbf {\bibinfo {volume} {10}},\ \bibinfo
  {pages} {031045} (\bibinfo {year} {2020})}\BibitemShut {NoStop}%
\bibitem [{\citenamefont {Franz}\ \emph {et~al.}(2025)\citenamefont {Franz},
  \citenamefont {Lupo}, \citenamefont {Nicoletti}, \citenamefont {Parisi},\
  and\ \citenamefont {Ricci-Tersenghi}}]{Franz2025-tp}%
  \BibitemOpen
  \bibfield  {author} {\bibinfo {author} {\bibfnamefont {S.}~\bibnamefont
  {Franz}}, \bibinfo {author} {\bibfnamefont {C.}~\bibnamefont {Lupo}},
  \bibinfo {author} {\bibfnamefont {F.}~\bibnamefont {Nicoletti}}, \bibinfo
  {author} {\bibfnamefont {G.}~\bibnamefont {Parisi}},\ and\ \bibinfo {author}
  {\bibfnamefont {F.}~\bibnamefont {Ricci-Tersenghi}},\ }\bibfield  {title}
  {\bibinfo {title} {Soft modes in vector spin glass models on sparse random
  graphs},\ }\href {https://doi.org/10.1103/physrevb.111.014203} {\bibfield
  {journal} {\bibinfo  {journal} {Phys. Rev. B}\ }\textbf {\bibinfo {volume}
  {111}},\ \bibinfo {pages} {014203} (\bibinfo {year} {2025})}\BibitemShut
  {NoStop}%
\bibitem [{\citenamefont {Vogel}\ \emph {et~al.}(2026)\citenamefont {Vogel},
  \citenamefont {Baumgärtel}, \citenamefont {Szamel},\ and\ \citenamefont
  {Fuchs}}]{Vogel2026-xe}%
  \BibitemOpen
  \bibfield  {author} {\bibinfo {author} {\bibfnamefont {F.}~\bibnamefont
  {Vogel}}, \bibinfo {author} {\bibfnamefont {P.}~\bibnamefont {Baumgärtel}},
  \bibinfo {author} {\bibfnamefont {G.}~\bibnamefont {Szamel}},\ and\ \bibinfo
  {author} {\bibfnamefont {M.}~\bibnamefont {Fuchs}},\ }\bibfield  {title}
  {\bibinfo {title} {On the euclidean random matrix model of vibrations in
  glass},\ }\href {https://doi.org/10.1038/s42005-026-02730-4} {\bibfield
  {journal} {\bibinfo  {journal} {Commun. Phys.}\ }\textbf {\bibinfo {volume}
  {9}},\ \bibinfo {pages} {223} (\bibinfo {year} {2026})}\BibitemShut {NoStop}%
\bibitem [{\citenamefont {Kittel}(2004)}]{Kittel2004}%
  \BibitemOpen
  \bibfield  {author} {\bibinfo {author} {\bibfnamefont {C.}~\bibnamefont
  {Kittel}},\ }\href@noop {} {\emph {\bibinfo {title} {Introduction to Solid
  State Physics}}},\ \bibinfo {edition} {8th}\ ed.\ (\bibinfo  {publisher}
  {John Wiley \& Sons},\ \bibinfo {address} {Hoboken, NJ},\ \bibinfo {year}
  {2004})\BibitemShut {NoStop}%
\bibitem [{\citenamefont {Sette}\ \emph {et~al.}(1998)\citenamefont {Sette},
  \citenamefont {Krisch}, \citenamefont {Masciovecchio}, \citenamefont
  {Ruocco},\ and\ \citenamefont {Monaco}}]{Sette1998-gc}%
  \BibitemOpen
  \bibfield  {author} {\bibinfo {author} {\bibfnamefont {F.}~\bibnamefont
  {Sette}}, \bibinfo {author} {\bibfnamefont {M.~H.}\ \bibnamefont {Krisch}},
  \bibinfo {author} {\bibfnamefont {C.}~\bibnamefont {Masciovecchio}}, \bibinfo
  {author} {\bibfnamefont {G.}~\bibnamefont {Ruocco}},\ and\ \bibinfo {author}
  {\bibfnamefont {G.}~\bibnamefont {Monaco}},\ }\bibfield  {title} {\bibinfo
  {title} {Dynamics of glasses and glass-forming liquids studied by inelastic
  {X}-ray scattering},\ }\href {https://doi.org/10.1126/science.280.5369.1550}
  {\bibfield  {journal} {\bibinfo  {journal} {Science}\ }\textbf {\bibinfo
  {volume} {280}},\ \bibinfo {pages} {1550} (\bibinfo {year}
  {1998})}\BibitemShut {NoStop}%
\bibitem [{\citenamefont {Grigera}\ \emph {et~al.}(2003)\citenamefont
  {Grigera}, \citenamefont {Martín-Mayor}, \citenamefont {Parisi},\ and\
  \citenamefont {Verrocchio}}]{Grigera2003-ym}%
  \BibitemOpen
  \bibfield  {author} {\bibinfo {author} {\bibfnamefont {T.~S.}\ \bibnamefont
  {Grigera}}, \bibinfo {author} {\bibfnamefont {V.}~\bibnamefont
  {Martín-Mayor}}, \bibinfo {author} {\bibfnamefont {G.}~\bibnamefont
  {Parisi}},\ and\ \bibinfo {author} {\bibfnamefont {P.}~\bibnamefont
  {Verrocchio}},\ }\bibfield  {title} {\bibinfo {title} {Phonon interpretation
  of the 'boson peak' in supercooled liquids},\ }\href
  {https://doi.org/10.1038/nature01475} {\bibfield  {journal} {\bibinfo
  {journal} {Nature}\ }\textbf {\bibinfo {volume} {422}},\ \bibinfo {pages}
  {289} (\bibinfo {year} {2003})}\BibitemShut {NoStop}%
\bibitem [{\citenamefont {Parisi}(2003)}]{Parisi2003-ex}%
  \BibitemOpen
  \bibfield  {author} {\bibinfo {author} {\bibfnamefont {G.}~\bibnamefont
  {Parisi}},\ }\bibfield  {title} {\bibinfo {title} {On the origin of the boson
  peak},\ }\href {https://doi.org/10.1088/0953-8984/15/11/302} {\bibfield
  {journal} {\bibinfo  {journal} {J. Phys. Condens. Matter}\ }\textbf {\bibinfo
  {volume} {15}},\ \bibinfo {pages} {S765} (\bibinfo {year}
  {2003})}\BibitemShut {NoStop}%
\bibitem [{\citenamefont {Leonforte}\ \emph {et~al.}(2005)\citenamefont
  {Leonforte}, \citenamefont {Boissière}, \citenamefont {Tanguy},
  \citenamefont {Wittmer},\ and\ \citenamefont {Barrat}}]{Leonforte2005-nm}%
  \BibitemOpen
  \bibfield  {author} {\bibinfo {author} {\bibfnamefont {F.}~\bibnamefont
  {Leonforte}}, \bibinfo {author} {\bibfnamefont {R.}~\bibnamefont
  {Boissière}}, \bibinfo {author} {\bibfnamefont {A.}~\bibnamefont {Tanguy}},
  \bibinfo {author} {\bibfnamefont {J.~P.}\ \bibnamefont {Wittmer}},\ and\
  \bibinfo {author} {\bibfnamefont {J.-L.}\ \bibnamefont {Barrat}},\ }\bibfield
   {title} {\bibinfo {title} {Continuum limit of amorphous elastic bodies.
  {III}. three-dimensional systems},\ }\href
  {https://doi.org/10.1103/physrevb.72.224206} {\bibfield  {journal} {\bibinfo
  {journal} {Phys. Rev. B}\ }\textbf {\bibinfo
  {volume} {72}},\ \bibinfo {pages} {224206} (\bibinfo {year}
  {2005})}\BibitemShut {NoStop}%
\bibitem [{\citenamefont {Schober}\ and\ \citenamefont
  {Laird}(1991)}]{Schober1991-in}%
  \BibitemOpen
  \bibfield  {author} {\bibinfo {author} {\bibfnamefont {H.~R.}\ \bibnamefont
  {Schober}}\ and\ \bibinfo {author} {\bibfnamefont {B.~B.}\ \bibnamefont
  {Laird}},\ }\bibfield  {title} {\bibinfo {title} {Localized low-frequency
  vibrational modes in glasses},\ }\href
  {https://doi.org/10.1103/physrevb.44.6746} {\bibfield  {journal} {\bibinfo
  {journal} {Phys. Rev. B}\ }\textbf {\bibinfo {volume} {44}},\
  \bibinfo {pages} {6746} (\bibinfo {year} {1991})}\BibitemShut {NoStop}%
\bibitem [{\citenamefont {Baity-Jesi}\ \emph {et~al.}(2015)\citenamefont
  {Baity-Jesi}, \citenamefont {Martín-Mayor}, \citenamefont {Parisi},\ and\
  \citenamefont {Perez-Gaviro}}]{Baity-Jesi2015-pz}%
  \BibitemOpen
  \bibfield  {author} {\bibinfo {author} {\bibfnamefont {M.}~\bibnamefont
  {Baity-Jesi}}, \bibinfo {author} {\bibfnamefont {V.}~\bibnamefont
  {Martín-Mayor}}, \bibinfo {author} {\bibfnamefont {G.}~\bibnamefont
  {Parisi}},\ and\ \bibinfo {author} {\bibfnamefont {S.}~\bibnamefont
  {Perez-Gaviro}},\ }\bibfield  {title} {\bibinfo {title} {Soft modes,
  localization, and two-level systems in spin glasses},\ }\href
  {https://doi.org/10.1103/PhysRevLett.115.267205} {\bibfield  {journal}
  {\bibinfo  {journal} {Phys. Rev. Lett.}\ }\textbf {\bibinfo {volume} {115}},\
  \bibinfo {pages} {267205} (\bibinfo {year} {2015})}\BibitemShut {NoStop}%
\bibitem [{\citenamefont {Lerner}\ \emph {et~al.}(2016)\citenamefont {Lerner},
  \citenamefont {Düring},\ and\ \citenamefont {Bouchbinder}}]{Lerner2016-ra}%
  \BibitemOpen
  \bibfield  {author} {\bibinfo {author} {\bibfnamefont {E.}~\bibnamefont
  {Lerner}}, \bibinfo {author} {\bibfnamefont {G.}~\bibnamefont {Düring}},\
  and\ \bibinfo {author} {\bibfnamefont {E.}~\bibnamefont {Bouchbinder}},\
  }\bibfield  {title} {\bibinfo {title} {Statistics and properties of
  low-frequency vibrational modes in structural glasses},\ }\href
  {https://doi.org/10.1103/PhysRevLett.117.035501} {\bibfield  {journal}
  {\bibinfo  {journal} {Phys. Rev. Lett.}\ }\textbf {\bibinfo {volume} {117}},\
  \bibinfo {pages} {035501} (\bibinfo {year} {2016})}\BibitemShut {NoStop}%
\bibitem [{\citenamefont {Mizuno}\ \emph {et~al.}(2017)\citenamefont {Mizuno},
  \citenamefont {Shiba},\ and\ \citenamefont {Ikeda}}]{Mizuno2017-sc}%
  \BibitemOpen
  \bibfield  {author} {\bibinfo {author} {\bibfnamefont {H.}~\bibnamefont
  {Mizuno}}, \bibinfo {author} {\bibfnamefont {H.}~\bibnamefont {Shiba}},\ and\
  \bibinfo {author} {\bibfnamefont {A.}~\bibnamefont {Ikeda}},\ }\bibfield
  {title} {\bibinfo {title} {Continuum limit of the vibrational properties of
  amorphous solids},\ }\href {https://doi.org/10.1073/pnas.1709015114}
  {\bibfield  {journal} {\bibinfo  {journal} {Proc. Natl. Acad. Sci. U. S. A.}\
  }\textbf {\bibinfo {volume} {114}},\ \bibinfo {pages} {E9767} (\bibinfo
  {year} {2017})}\BibitemShut {NoStop}%
\bibitem [{\citenamefont {Angelani}\ \emph {et~al.}(2018)\citenamefont
  {Angelani}, \citenamefont {Paoluzzi}, \citenamefont {Parisi},\ and\
  \citenamefont {Ruocco}}]{Angelani2018-vt}%
  \BibitemOpen
  \bibfield  {author} {\bibinfo {author} {\bibfnamefont {L.}~\bibnamefont
  {Angelani}}, \bibinfo {author} {\bibfnamefont {M.}~\bibnamefont {Paoluzzi}},
  \bibinfo {author} {\bibfnamefont {G.}~\bibnamefont {Parisi}},\ and\ \bibinfo
  {author} {\bibfnamefont {G.}~\bibnamefont {Ruocco}},\ }\bibfield  {title}
  {\bibinfo {title} {Probing the non-debye low-frequency excitations in glasses
  through random pinning},\ }\href {https://doi.org/10.1073/pnas.1805024115}
  {\bibfield  {journal} {\bibinfo  {journal} {Proc. Natl. Acad. Sci. U. S. A.}\
  }\textbf {\bibinfo {volume} {115}},\ \bibinfo {pages} {8700} (\bibinfo {year}
  {2018})}\BibitemShut {NoStop}%
\bibitem [{\citenamefont {Wang}\ \emph {et~al.}(2019)\citenamefont {Wang},
  \citenamefont {Ninarello}, \citenamefont {Guan}, \citenamefont {Berthier},
  \citenamefont {Szamel},\ and\ \citenamefont {Flenner}}]{Wang2019-jq}%
  \BibitemOpen
  \bibfield  {author} {\bibinfo {author} {\bibfnamefont {L.}~\bibnamefont
  {Wang}}, \bibinfo {author} {\bibfnamefont {A.}~\bibnamefont {Ninarello}},
  \bibinfo {author} {\bibfnamefont {P.}~\bibnamefont {Guan}}, \bibinfo {author}
  {\bibfnamefont {L.}~\bibnamefont {Berthier}}, \bibinfo {author}
  {\bibfnamefont {G.}~\bibnamefont {Szamel}},\ and\ \bibinfo {author}
  {\bibfnamefont {E.}~\bibnamefont {Flenner}},\ }\bibfield  {title} {\bibinfo
  {title} {Low-frequency vibrational modes of stable glasses},\ }\href
  {https://doi.org/10.1038/s41467-018-07978-1} {\bibfield  {journal} {\bibinfo
  {journal} {Nat. Commun.}\ }\textbf {\bibinfo {volume} {10}},\ \bibinfo
  {pages} {26} (\bibinfo {year} {2019})}\BibitemShut {NoStop}%
\bibitem [{\citenamefont {Bonfanti}\ \emph {et~al.}(2020)\citenamefont
  {Bonfanti}, \citenamefont {Guerra}, \citenamefont {Mondal}, \citenamefont
  {Procaccia},\ and\ \citenamefont {Zapperi}}]{Bonfanti2020-nv}%
  \BibitemOpen
  \bibfield  {author} {\bibinfo {author} {\bibfnamefont {S.}~\bibnamefont
  {Bonfanti}}, \bibinfo {author} {\bibfnamefont {R.}~\bibnamefont {Guerra}},
  \bibinfo {author} {\bibfnamefont {C.}~\bibnamefont {Mondal}}, \bibinfo
  {author} {\bibfnamefont {I.}~\bibnamefont {Procaccia}},\ and\ \bibinfo
  {author} {\bibfnamefont {S.}~\bibnamefont {Zapperi}},\ }\bibfield  {title}
  {\bibinfo {title} {Universal low-frequency vibrational modes in silica
  glasses},\ }\href {https://doi.org/10.1103/PhysRevLett.125.085501} {\bibfield
   {journal} {\bibinfo  {journal} {Phys. Rev. Lett.}\ }\textbf {\bibinfo
  {volume} {125}},\ \bibinfo {pages} {085501} (\bibinfo {year}
  {2020})}\BibitemShut {NoStop}%
\bibitem [{\citenamefont {Lerner}\ and\ \citenamefont
  {Bouchbinder}(2021)}]{Lerner2021-en}%
  \BibitemOpen
  \bibfield  {author} {\bibinfo {author} {\bibfnamefont {E.}~\bibnamefont
  {Lerner}}\ and\ \bibinfo {author} {\bibfnamefont {E.}~\bibnamefont
  {Bouchbinder}},\ }\bibfield  {title} {\bibinfo {title} {Low-energy
  quasilocalized excitations in structural glasses},\ }\href
  {https://doi.org/10.1063/5.0069477} {\bibfield  {journal} {\bibinfo
  {journal} {J. Chem. Phys.}\ }\textbf {\bibinfo {volume} {155}},\ \bibinfo
  {pages} {200901} (\bibinfo {year} {2021})}\BibitemShut {NoStop}%
\bibitem [{\citenamefont {Wang}\ \emph {et~al.}(2025)\citenamefont {Wang},
  \citenamefont {Xu}, \citenamefont {Zhang}, \citenamefont {Nie}, \citenamefont
  {Tong},\ and\ \citenamefont {Xu}}]{Wang2025-so}%
  \BibitemOpen
  \bibfield  {author} {\bibinfo {author} {\bibfnamefont {L.}~\bibnamefont
  {Wang}}, \bibinfo {author} {\bibfnamefont {D.}~\bibnamefont {Xu}}, \bibinfo
  {author} {\bibfnamefont {S.}~\bibnamefont {Zhang}}, \bibinfo {author}
  {\bibfnamefont {Y.}~\bibnamefont {Nie}}, \bibinfo {author} {\bibfnamefont
  {H.}~\bibnamefont {Tong}},\ and\ \bibinfo {author} {\bibfnamefont
  {N.}~\bibnamefont {Xu}},\ }\bibfield  {title} {\bibinfo {title}
  {Low-frequency non-phononic vibrations of amorphous solids},\ }\href
  {https://doi.org/10.1088/1361-6633/ae0e34} {\bibfield  {journal} {\bibinfo
  {journal} {Rep. Prog. Phys.}\ }\textbf {\bibinfo {volume} {88}},\ \bibinfo
  {pages} {104601} (\bibinfo {year} {2025})}\BibitemShut {NoStop}%
\bibitem [{\citenamefont {Ramos}(2022)}]{Ramos2022-nf}%
  \BibitemOpen
  \bibinfo {editor} {\bibfnamefont {M.~A.}\ \bibnamefont {Ramos}},\ ed.,\ \href
  {https://doi.org/10.1142/q0371} {\emph {\bibinfo {title} {Low-temperature
  thermal and vibrational properties of disordered solids: A half-century of
  universal ``anomalies'' of glasses}}}\ (\bibinfo  {publisher} {World
  Scientific Europe},\ \bibinfo {address} {London, England},\ \bibinfo {year}
  {2022})\BibitemShut {NoStop}%
\bibitem [{\citenamefont {Buchenau}\ \emph {et~al.}(1984)\citenamefont
  {Buchenau}, \citenamefont {Nücker},\ and\ \citenamefont
  {Dianoux}}]{Buchenau1984-dy}%
  \BibitemOpen
  \bibfield  {author} {\bibinfo {author} {\bibfnamefont {U.}~\bibnamefont
  {Buchenau}}, \bibinfo {author} {\bibfnamefont {N.}~\bibnamefont {Nücker}},\
  and\ \bibinfo {author} {\bibfnamefont {A.~J.}\ \bibnamefont {Dianoux}},\
  }\bibfield  {title} {\bibinfo {title} {Neutron scattering study of the
  low-frequency vibrations in vitreous silica},\ }\href
  {https://doi.org/10.1103/PhysRevLett.53.2316} {\bibfield  {journal} {\bibinfo
   {journal} {Phys. Rev. Lett.}\ }\textbf {\bibinfo {volume} {53}},\ \bibinfo
  {pages} {2316} (\bibinfo {year} {1984})}\BibitemShut {NoStop}%
\bibitem [{\citenamefont {Malinovsky}\ and\ \citenamefont
  {Sokolov}(1986)}]{Malinovsky1986-vb}%
  \BibitemOpen
  \bibfield  {author} {\bibinfo {author} {\bibfnamefont {V.~K.}\ \bibnamefont
  {Malinovsky}}\ and\ \bibinfo {author} {\bibfnamefont {A.~P.}\ \bibnamefont
  {Sokolov}},\ }\bibfield  {title} {\bibinfo {title} {The nature of boson peak
  in raman scattering in glasses},\ }\href
  {https://doi.org/10.1016/0038-1098(86)90854-9} {\bibfield  {journal}
  {\bibinfo  {journal} {Solid State Commun.}\ }\textbf {\bibinfo {volume}
  {57}},\ \bibinfo {pages} {757} (\bibinfo {year} {1986})}\BibitemShut
  {NoStop}%
\bibitem [{\citenamefont {Schirmacher}\ \emph {et~al.}(1998)\citenamefont
  {Schirmacher}, \citenamefont {Diezemann},\ and\ \citenamefont
  {Ganter}}]{Schirmacher1998-fa}%
  \BibitemOpen
  \bibfield  {author} {\bibinfo {author} {\bibfnamefont {W.}~\bibnamefont
  {Schirmacher}}, \bibinfo {author} {\bibfnamefont {G.}~\bibnamefont
  {Diezemann}},\ and\ \bibinfo {author} {\bibfnamefont {C.}~\bibnamefont
  {Ganter}},\ }\bibfield  {title} {\bibinfo {title} {Harmonic vibrational
  excitations in disordered solids and the “boson peak”},\ }\href
  {https://doi.org/10.1103/physrevlett.81.136} {\bibfield  {journal} {\bibinfo
  {journal} {Phys. Rev. Lett.}\ }\textbf {\bibinfo {volume} {81}},\ \bibinfo
  {pages} {136} (\bibinfo {year} {1998})}\BibitemShut {NoStop}%
\bibitem [{\citenamefont {Nakayama}(2002)}]{Nakayama2002-ve}%
  \BibitemOpen
  \bibfield  {author} {\bibinfo {author} {\bibfnamefont {T.}~\bibnamefont
  {Nakayama}},\ }\bibfield  {title} {\bibinfo {title} {Boson peak and terahertz
  frequency dynamics of vitreous silica},\ }\href
  {https://doi.org/10.1088/0034-4885/65/8/203} {\bibfield  {journal} {\bibinfo
  {journal} {Rep. Prog. Phys.}\ }\textbf {\bibinfo {volume} {65}},\ \bibinfo
  {pages} {1195} (\bibinfo {year} {2002})}\BibitemShut {NoStop}%
\bibitem [{\citenamefont {Chumakov}\ \emph {et~al.}(2004)\citenamefont
  {Chumakov}, \citenamefont {Sergueev}, \citenamefont {van Bürck},
  \citenamefont {Schirmacher}, \citenamefont {Asthalter}, \citenamefont
  {Rüffer}, \citenamefont {Leupold},\ and\ \citenamefont
  {Petry}}]{Chumakov2004-gb}%
  \BibitemOpen
  \bibfield  {author} {\bibinfo {author} {\bibfnamefont {A.~I.}\ \bibnamefont
  {Chumakov}}, \bibinfo {author} {\bibfnamefont {I.}~\bibnamefont {Sergueev}},
  \bibinfo {author} {\bibfnamefont {U.}~\bibnamefont {van Bürck}}, \bibinfo
  {author} {\bibfnamefont {W.}~\bibnamefont {Schirmacher}}, \bibinfo {author}
  {\bibfnamefont {T.}~\bibnamefont {Asthalter}}, \bibinfo {author}
  {\bibfnamefont {R.}~\bibnamefont {Rüffer}}, \bibinfo {author} {\bibfnamefont
  {O.}~\bibnamefont {Leupold}},\ and\ \bibinfo {author} {\bibfnamefont
  {W.}~\bibnamefont {Petry}},\ }\bibfield  {title} {\bibinfo {title}
  {Collective nature of the boson peak and universal transboson dynamics of
  glasses},\ }\href {https://doi.org/10.1103/PhysRevLett.92.245508} {\bibfield
  {journal} {\bibinfo  {journal} {Phys. Rev. Lett.}\ }\textbf {\bibinfo
  {volume} {92}},\ \bibinfo {pages} {245508} (\bibinfo {year}
  {2004})}\BibitemShut {NoStop}%
\bibitem [{\citenamefont {González-Jiménez}\ \emph
  {et~al.}(2023)\citenamefont {González-Jiménez}, \citenamefont {Barnard},
  \citenamefont {Russell}, \citenamefont {Tukachev}, \citenamefont {Javornik},
  \citenamefont {Hayes}, \citenamefont {Farrell}, \citenamefont {Guinane},
  \citenamefont {Senn}, \citenamefont {Smith}, \citenamefont {Wilding},
  \citenamefont {Mali}, \citenamefont {Nakano}, \citenamefont {Miyazaki},
  \citenamefont {McMillan}, \citenamefont {Sosso},\ and\ \citenamefont
  {Wynne}}]{Gonzalez-Jimenez2023-un}%
  \BibitemOpen
  \bibfield  {author} {\bibinfo {author} {\bibfnamefont {M.}~\bibnamefont
  {González-Jiménez}}, \bibinfo {author} {\bibfnamefont {T.}~\bibnamefont
  {Barnard}}, \bibinfo {author} {\bibfnamefont {B.~A.}\ \bibnamefont
  {Russell}}, \bibinfo {author} {\bibfnamefont {N.~V.}\ \bibnamefont
  {Tukachev}}, \bibinfo {author} {\bibfnamefont {U.}~\bibnamefont {Javornik}},
  \bibinfo {author} {\bibfnamefont {L.-A.}\ \bibnamefont {Hayes}}, \bibinfo
  {author} {\bibfnamefont {A.~J.}\ \bibnamefont {Farrell}}, \bibinfo {author}
  {\bibfnamefont {S.}~\bibnamefont {Guinane}}, \bibinfo {author} {\bibfnamefont
  {H.~M.}\ \bibnamefont {Senn}}, \bibinfo {author} {\bibfnamefont {A.~J.}\
  \bibnamefont {Smith}}, \bibinfo {author} {\bibfnamefont {M.}~\bibnamefont
  {Wilding}}, \bibinfo {author} {\bibfnamefont {G.}~\bibnamefont {Mali}},
  \bibinfo {author} {\bibfnamefont {M.}~\bibnamefont {Nakano}}, \bibinfo
  {author} {\bibfnamefont {Y.}~\bibnamefont {Miyazaki}}, \bibinfo {author}
  {\bibfnamefont {P.}~\bibnamefont {McMillan}}, \bibinfo {author}
  {\bibfnamefont {G.~C.}\ \bibnamefont {Sosso}},\ and\ \bibinfo {author}
  {\bibfnamefont {K.}~\bibnamefont {Wynne}},\ }\bibfield  {title} {\bibinfo
  {title} {Understanding the emergence of the boson peak in molecular
  glasses},\ }\href {https://doi.org/10.1038/s41467-023-35878-6} {\bibfield
  {journal} {\bibinfo  {journal} {Nat. Commun.}\ }\textbf {\bibinfo {volume}
  {14}},\ \bibinfo {pages} {215} (\bibinfo {year} {2023})}\BibitemShut
  {NoStop}%
\bibitem [{\citenamefont {Festi}\ \emph {et~al.}(2026)\citenamefont {Festi},
  \citenamefont {Alfinelli}, \citenamefont {Bessas}, \citenamefont
  {Caporaletti}, \citenamefont {Chumakov}, \citenamefont {Moratalla},
  \citenamefont {Ramos}, \citenamefont {Rodríguez-López}, \citenamefont
  {Rodríguez-Tinoco}, \citenamefont {Rodríguez-Viejo},\ and\ \citenamefont
  {Baldi}}]{Festi2026-zu}%
  \BibitemOpen
  \bibfield  {author} {\bibinfo {author} {\bibfnamefont {I.}~\bibnamefont
  {Festi}}, \bibinfo {author} {\bibfnamefont {E.}~\bibnamefont {Alfinelli}},
  \bibinfo {author} {\bibfnamefont {D.}~\bibnamefont {Bessas}}, \bibinfo
  {author} {\bibfnamefont {F.}~\bibnamefont {Caporaletti}}, \bibinfo {author}
  {\bibfnamefont {A.~I.}\ \bibnamefont {Chumakov}}, \bibinfo {author}
  {\bibfnamefont {M.}~\bibnamefont {Moratalla}}, \bibinfo {author}
  {\bibfnamefont {M.~A.}\ \bibnamefont {Ramos}}, \bibinfo {author}
  {\bibfnamefont {M.}~\bibnamefont {Rodríguez-López}}, \bibinfo {author}
  {\bibfnamefont {C.}~\bibnamefont {Rodríguez-Tinoco}}, \bibinfo {author}
  {\bibfnamefont {J.}~\bibnamefont {Rodríguez-Viejo}},\ and\ \bibinfo {author}
  {\bibfnamefont {G.}~\bibnamefont {Baldi}},\ }\bibfield  {title} {\bibinfo
  {title} {Effect of glass stability on the low frequency vibrations of vapor
  deposited glasses},\ }\href {https://doi.org/10.1103/311v-1ftn} {\bibfield
  {journal} {\bibinfo  {journal} {Phys. Rev. X}\ }\textbf {\bibinfo {volume}
  {16}},\ \bibinfo {pages} {021021} (\bibinfo {year} {2026})}\BibitemShut
  {NoStop}%
\bibitem [{\citenamefont {Lerner}\ and\ \citenamefont
  {Bouchbinder}(2023)}]{Lerner2023-dp}%
  \BibitemOpen
  \bibfield  {author} {\bibinfo {author} {\bibfnamefont {E.}~\bibnamefont
  {Lerner}}\ and\ \bibinfo {author} {\bibfnamefont {E.}~\bibnamefont
  {Bouchbinder}},\ }\bibfield  {title} {\bibinfo {title} {Boson-peak
  vibrational modes in glasses feature hybridized phononic and quasilocalized
  excitations},\ }\href {https://doi.org/10.1063/5.0147889} {\bibfield
  {journal} {\bibinfo  {journal} {J. Chem. Phys.}\ }\textbf {\bibinfo {volume}
  {158}},\ \bibinfo {pages} {194503} (\bibinfo {year} {2023})}\BibitemShut
  {NoStop}%
\bibitem [{\citenamefont {Moriel}\ \emph {et~al.}(2024)\citenamefont {Moriel},
  \citenamefont {Lerner},\ and\ \citenamefont {Bouchbinder}}]{Moriel2024-wp}%
  \BibitemOpen
  \bibfield  {author} {\bibinfo {author} {\bibfnamefont {A.}~\bibnamefont
  {Moriel}}, \bibinfo {author} {\bibfnamefont {E.}~\bibnamefont {Lerner}},\
  and\ \bibinfo {author} {\bibfnamefont {E.}~\bibnamefont {Bouchbinder}},\
  }\bibfield  {title} {\bibinfo {title} {Boson peak in the vibrational spectra
  of glasses},\ }\href {https://doi.org/10.1103/physrevresearch.6.023053}
  {\bibfield  {journal} {\bibinfo  {journal} {Phys. Rev. Res.}\ }\textbf
  {\bibinfo {volume} {6}},\ \bibinfo {pages} {023053} (\bibinfo {year}
  {2024})}\BibitemShut {NoStop}%
\bibitem [{\citenamefont {Bouchbinder}\ \emph {et~al.}(2021)\citenamefont
  {Bouchbinder}, \citenamefont {Lerner}, \citenamefont {Rainone}, \citenamefont
  {Urbani},\ and\ \citenamefont {Zamponi}}]{Bouchbinder2021-dh}%
  \BibitemOpen
  \bibfield  {author} {\bibinfo {author} {\bibfnamefont {E.}~\bibnamefont
  {Bouchbinder}}, \bibinfo {author} {\bibfnamefont {E.}~\bibnamefont {Lerner}},
  \bibinfo {author} {\bibfnamefont {C.}~\bibnamefont {Rainone}}, \bibinfo
  {author} {\bibfnamefont {P.}~\bibnamefont {Urbani}},\ and\ \bibinfo {author}
  {\bibfnamefont {F.}~\bibnamefont {Zamponi}},\ }\bibfield  {title} {\bibinfo
  {title} {Low-frequency vibrational spectrum of mean-field disordered
  systems},\ }\href {https://doi.org/10.1103/PhysRevB.103.174202} {\bibfield
  {journal} {\bibinfo  {journal} {Phys. Rev. B}\ }\textbf {\bibinfo {volume}
  {103}},\ \bibinfo {pages} {174202} (\bibinfo {year} {2021})}\BibitemShut
  {NoStop}%
\bibitem [{\citenamefont {Rainone}\ \emph {et~al.}(2021)\citenamefont
  {Rainone}, \citenamefont {Urbani}, \citenamefont {Zamponi}, \citenamefont
  {Lerner},\ and\ \citenamefont {Bouchbinder}}]{Rainone2021-am}%
  \BibitemOpen
  \bibfield  {author} {\bibinfo {author} {\bibfnamefont {C.}~\bibnamefont
  {Rainone}}, \bibinfo {author} {\bibfnamefont {P.}~\bibnamefont {Urbani}},
  \bibinfo {author} {\bibfnamefont {F.}~\bibnamefont {Zamponi}}, \bibinfo
  {author} {\bibfnamefont {E.}~\bibnamefont {Lerner}},\ and\ \bibinfo {author}
  {\bibfnamefont {E.}~\bibnamefont {Bouchbinder}},\ }\bibfield  {title}
  {\bibinfo {title} {Mean-field model of interacting quasilocalized excitations
  in glasses},\ }\href {https://doi.org/10.21468/scipostphyscore.4.2.008}
  {\bibfield  {journal} {\bibinfo  {journal} {SciPost Phys. Core}\ }\textbf
  {\bibinfo {volume} {4}},\ \bibinfo {pages} {008} (\bibinfo {year}
  {2021})}\BibitemShut {NoStop}%
\bibitem [{\citenamefont {Suda}\ \emph {et~al.}(2026)\citenamefont {Suda},
  \citenamefont {Lerner},\ and\ \citenamefont {Bouchbinder}}]{Suda2026-cl}%
  \BibitemOpen
  \bibfield  {author} {\bibinfo {author} {\bibfnamefont {M.}~\bibnamefont
  {Suda}}, \bibinfo {author} {\bibfnamefont {E.}~\bibnamefont {Lerner}},\ and\
  \bibinfo {author} {\bibfnamefont {E.}~\bibnamefont {Bouchbinder}},\
  }\bibfield  {title} {\bibinfo {title} {Local micromechanics in a mean-field
  model of glasses reveal key properties of its nonequilibrium
  replica-symmetry-breaking phase},\ }\href {https://doi.org/10.1103/wtds-jlcz}
  {\bibfield  {journal} {\bibinfo  {journal} {Phys. Rev. B}\ }\textbf {\bibinfo
  {volume} {114}},\ \bibinfo {pages} {134113} (\bibinfo {year}
  {2026})}\BibitemShut {NoStop}%
\bibitem [{\citenamefont {Suda}\ \emph {et~al.}(2025)\citenamefont {Suda},
  \citenamefont {Lerner},\ and\ \citenamefont {Bouchbinder}}]{Suda2025-kl}%
  \BibitemOpen
  \bibfield  {author} {\bibinfo {author} {\bibfnamefont {M.}~\bibnamefont
  {Suda}}, \bibinfo {author} {\bibfnamefont {E.}~\bibnamefont {Lerner}},\ and\
  \bibinfo {author} {\bibfnamefont {E.}~\bibnamefont {Bouchbinder}},\
  }\bibfield  {title} {\bibinfo {title} {Yielding and memory in a driven
  mean-field model of glasses},\ }\href {https://doi.org/10.1103/vpmn-sw7n}
  {\bibfield  {journal} {\bibinfo  {journal} {Phys. Rev. Lett.}\ }\textbf
  {\bibinfo {volume} {135}},\ \bibinfo {pages} {258201} (\bibinfo {year}
  {2025})}\BibitemShut {NoStop}%
\bibitem [{\citenamefont {Gartner}\ and\ \citenamefont
  {Lerner}(2016{\natexlab{a}})}]{Gartner2016-vw}%
  \BibitemOpen
  \bibfield  {author} {\bibinfo {author} {\bibfnamefont {L.}~\bibnamefont
  {Gartner}}\ and\ \bibinfo {author} {\bibfnamefont {E.}~\bibnamefont
  {Lerner}},\ }\bibfield  {title} {\bibinfo {title} {Nonlinear plastic modes in
  disordered solids},\ }\href {https://doi.org/10.1103/PhysRevE.93.011001}
  {\bibfield  {journal} {\bibinfo  {journal} {Phys. Rev. E}\ }\textbf
  {\bibinfo {volume} {93}},\ \bibinfo {pages} {011001} (\bibinfo {year}
  {2016}{\natexlab{a}})}\BibitemShut {NoStop}%
\bibitem [{\citenamefont {Gartner}\ and\ \citenamefont
  {Lerner}(2016{\natexlab{b}})}]{Gartner2016-lc}%
  \BibitemOpen
  \bibfield  {author} {\bibinfo {author} {\bibfnamefont {L.}~\bibnamefont
  {Gartner}}\ and\ \bibinfo {author} {\bibfnamefont {E.}~\bibnamefont
  {Lerner}},\ }\bibfield  {title} {\bibinfo {title} {Nonlinear modes
  disentangle glassy and goldstone modes in structural glasses},\ }\href
  {https://doi.org/10.21468/scipostphys.1.2.016} {\bibfield  {journal}
  {\bibinfo  {journal} {SciPost Phys.}\ }\textbf {\bibinfo {volume} {1}},\
  \bibinfo {pages} {016} (\bibinfo {year} {2016}{\natexlab{b}})}\BibitemShut
  {NoStop}%
\bibitem [{\citenamefont {Lerner}(2016)}]{Lerner2016-ep}%
  \BibitemOpen
  \bibfield  {author} {\bibinfo {author} {\bibfnamefont {E.}~\bibnamefont
  {Lerner}},\ }\bibfield  {title} {\bibinfo {title} {Micromechanics of
  nonlinear plastic modes},\ }\href
  {https://doi.org/10.1103/PhysRevE.93.053004} {\bibfield  {journal} {\bibinfo
  {journal} {Phys. Rev. E}\ }\textbf {\bibinfo {volume} {93}},\ \bibinfo
  {pages} {053004} (\bibinfo {year} {2016})}\BibitemShut {NoStop}%
\bibitem [{\citenamefont {Kapteijns}\ \emph {et~al.}(2020)\citenamefont
  {Kapteijns}, \citenamefont {Richard},\ and\ \citenamefont
  {Lerner}}]{Kapteijns2020-fu}%
  \BibitemOpen
  \bibfield  {author} {\bibinfo {author} {\bibfnamefont {G.}~\bibnamefont
  {Kapteijns}}, \bibinfo {author} {\bibfnamefont {D.}~\bibnamefont {Richard}},\
  and\ \bibinfo {author} {\bibfnamefont {E.}~\bibnamefont {Lerner}},\
  }\bibfield  {title} {\bibinfo {title} {Nonlinear quasilocalized excitations
  in glasses: True representatives of soft spots},\ }\href
  {https://doi.org/10.1103/PhysRevE.101.032130} {\bibfield  {journal} {\bibinfo
   {journal} {Phys. Rev. E}\ }\textbf {\bibinfo {volume} {101}},\ \bibinfo
  {pages} {032130} (\bibinfo {year} {2020})}\BibitemShut {NoStop}%
\bibitem [{\citenamefont {Richard}\ \emph {et~al.}(2020)\citenamefont
  {Richard}, \citenamefont {Ozawa}, \citenamefont {Patinet}, \citenamefont
  {Stanifer}, \citenamefont {Shang}, \citenamefont {Ridout}, \citenamefont
  {Xu}, \citenamefont {Zhang}, \citenamefont {Morse}, \citenamefont {Barrat},
  \citenamefont {Berthier}, \citenamefont {Falk}, \citenamefont {Guan},
  \citenamefont {Liu}, \citenamefont {Martens}, \citenamefont {Sastry},
  \citenamefont {Vandembroucq}, \citenamefont {Lerner},\ and\ \citenamefont
  {Manning}}]{Richard2020-vr}%
  \BibitemOpen
  \bibfield  {author} {\bibinfo {author} {\bibfnamefont {D.}~\bibnamefont
  {Richard}}, \bibinfo {author} {\bibfnamefont {M.}~\bibnamefont {Ozawa}},
  \bibinfo {author} {\bibfnamefont {S.}~\bibnamefont {Patinet}}, \bibinfo
  {author} {\bibfnamefont {E.}~\bibnamefont {Stanifer}}, \bibinfo {author}
  {\bibfnamefont {B.}~\bibnamefont {Shang}}, \bibinfo {author} {\bibfnamefont
  {S.~A.}\ \bibnamefont {Ridout}}, \bibinfo {author} {\bibfnamefont
  {B.}~\bibnamefont {Xu}}, \bibinfo {author} {\bibfnamefont {G.}~\bibnamefont
  {Zhang}}, \bibinfo {author} {\bibfnamefont {P.~K.}\ \bibnamefont {Morse}},
  \bibinfo {author} {\bibfnamefont {J.-L.}\ \bibnamefont {Barrat}}, \bibinfo
  {author} {\bibfnamefont {L.}~\bibnamefont {Berthier}}, \bibinfo {author}
  {\bibfnamefont {M.~L.}\ \bibnamefont {Falk}}, \bibinfo {author}
  {\bibfnamefont {P.}~\bibnamefont {Guan}}, \bibinfo {author} {\bibfnamefont
  {A.~J.}\ \bibnamefont {Liu}}, \bibinfo {author} {\bibfnamefont
  {K.}~\bibnamefont {Martens}}, \bibinfo {author} {\bibfnamefont
  {S.}~\bibnamefont {Sastry}}, \bibinfo {author} {\bibfnamefont
  {D.}~\bibnamefont {Vandembroucq}}, \bibinfo {author} {\bibfnamefont
  {E.}~\bibnamefont {Lerner}},\ and\ \bibinfo {author} {\bibfnamefont {M.~L.}\
  \bibnamefont {Manning}},\ }\bibfield  {title} {\bibinfo {title} {Predicting
  plasticity in disordered solids from structural indicators},\ }\href
  {https://doi.org/10.1103/physrevmaterials.4.113609} {\bibfield  {journal}
  {\bibinfo  {journal} {Phys. Rev. Mater.}\ }\textbf {\bibinfo {volume} {4}},\
  \bibinfo {pages} {113609} (\bibinfo {year} {2020})}\BibitemShut {NoStop}%
\bibitem [{\citenamefont {Anderson}\ \emph {et~al.}(1972)\citenamefont
  {Anderson}, \citenamefont {Halperin},\ and\ \citenamefont
  {Varma}}]{Anderson1972-bi}%
  \BibitemOpen
  \bibfield  {author} {\bibinfo {author} {\bibfnamefont {P.~W.}\ \bibnamefont
  {Anderson}}, \bibinfo {author} {\bibfnamefont {B.~I.}\ \bibnamefont
  {Halperin}},\ and\ \bibinfo {author} {\bibfnamefont {C.~M.}\ \bibnamefont
  {Varma}},\ }\bibfield  {title} {\bibinfo {title} {Anomalous low-temperature
  thermal properties of glasses and spin glasses},\ }\href
  {https://doi.org/10.1080/14786437208229210} {\bibfield  {journal} {\bibinfo
  {journal} {Philos. Mag.}\ }\textbf {\bibinfo {volume} {25}},\ \bibinfo
  {pages} {1} (\bibinfo {year} {1972})}\BibitemShut {NoStop}%
\bibitem [{\citenamefont {Phillips}(1987)}]{Phillips1987-ti}%
  \BibitemOpen
  \bibfield  {author} {\bibinfo {author} {\bibfnamefont {W.~A.}\ \bibnamefont
  {Phillips}},\ }\bibfield  {title} {\bibinfo {title} {Two-level states in
  glasses},\ }\href {https://doi.org/10.1088/0034-4885/50/12/003} {\bibfield
  {journal} {\bibinfo  {journal} {Rep. Prog. Phys.}\ }\textbf {\bibinfo
  {volume} {50}},\ \bibinfo {pages} {1657} (\bibinfo {year}
  {1987})}\BibitemShut {NoStop}%
\bibitem [{\citenamefont {Heuer}(2008)}]{Heuer2008-ou}%
  \BibitemOpen
  \bibfield  {author} {\bibinfo {author} {\bibfnamefont {A.}~\bibnamefont
  {Heuer}},\ }\bibfield  {title} {\bibinfo {title} {Exploring the potential
  energy landscape of glass-forming systems: from inherent structures via
  metabasins to macroscopic transport},\ }\href
  {https://doi.org/10.1088/0953-8984/20/37/373101} {\bibfield  {journal}
  {\bibinfo  {journal} {J. Phys. Condens. Matter}\ }\textbf {\bibinfo {volume}
  {20}},\ \bibinfo {pages} {373101} (\bibinfo {year} {2008})}\BibitemShut
  {NoStop}%
\bibitem [{\citenamefont {Falk}\ and\ \citenamefont
  {Langer}(2011)}]{Falk2011-nd}%
  \BibitemOpen
  \bibfield  {author} {\bibinfo {author} {\bibfnamefont {M.~L.}\ \bibnamefont
  {Falk}}\ and\ \bibinfo {author} {\bibfnamefont {J.~S.}\ \bibnamefont
  {Langer}},\ }\bibfield  {title} {\bibinfo {title} {Deformation and failure of
  amorphous, solidlike materials},\ }\href
  {https://doi.org/10.1146/annurev-conmatphys-062910-140452} {\bibfield
  {journal} {\bibinfo  {journal} {Annu. Rev. Condens. Matter Phys.}\ }\textbf
  {\bibinfo {volume} {2}},\ \bibinfo {pages} {353} (\bibinfo {year}
  {2011})}\BibitemShut {NoStop}%
\bibitem [{\citenamefont {Berthier}\ and\ \citenamefont
  {Reichman}(2023)}]{Berthier2023-cf}%
  \BibitemOpen
  \bibfield  {author} {\bibinfo {author} {\bibfnamefont {L.}~\bibnamefont
  {Berthier}}\ and\ \bibinfo {author} {\bibfnamefont {D.~R.}\ \bibnamefont
  {Reichman}},\ }\bibfield  {title} {\bibinfo {title} {Modern computational
  studies of the glass transition},\ }\href
  {https://doi.org/10.1038/s42254-022-00548-x} {\bibfield  {journal} {\bibinfo
  {journal} {Nat. Rev. Phys.}\ }\textbf {\bibinfo {volume} {5}},\ \bibinfo
  {pages} {102} (\bibinfo {year} {2023})}\BibitemShut {NoStop}%
\bibitem [{\citenamefont {Blaber}\ \emph {et~al.}(2026)\citenamefont {Blaber},
  \citenamefont {Bruns},\ and\ \citenamefont {Rottler}}]{Blaber2026-ic}%
  \BibitemOpen
  \bibfield  {author} {\bibinfo {author} {\bibfnamefont {S.}~\bibnamefont
  {Blaber}}, \bibinfo {author} {\bibfnamefont {D.}~\bibnamefont {Bruns}},\ and\
  \bibinfo {author} {\bibfnamefont {J.}~\bibnamefont {Rottler}},\ }\bibfield
  {title} {\bibinfo {title} {Connected network model for the mechanical loss of
  amorphous materials},\ }\href {https://doi.org/10.1073/pnas.2513161123}
  {\bibfield  {journal} {\bibinfo  {journal} {Proc. Natl. Acad. Sci. U. S. A.}\
  }\textbf {\bibinfo {volume} {123}},\ \bibinfo {pages} {e2513161123} (\bibinfo
  {year} {2026})}\BibitemShut {NoStop}%
\bibitem [{\citenamefont {Karpov}\ \emph {et~al.}(1983)\citenamefont {Karpov},
  \citenamefont {Klinger},\ and\ \citenamefont {Ignat'ev}}]{Karpov1983-ri}%
  \BibitemOpen
  \bibfield  {author} {\bibinfo {author} {\bibfnamefont {V.~G.}\ \bibnamefont
  {Karpov}}, \bibinfo {author} {\bibfnamefont {M.~I.}\ \bibnamefont
  {Klinger}},\ and\ \bibinfo {author} {\bibfnamefont {F.~N.}\ \bibnamefont
  {Ignat'ev}},\ }\bibfield  {title} {\bibinfo {title} {Theory of the
  low-temperature anomalies in the thermal properties of amorphous
  structures},\ }\href {https://jetp.ras.ru/cgi-bin/dn/e_057_02_0439.pdf}
  {\bibfield  {journal} {\bibinfo  {journal} {Sov. Phys., JETP}\ }\textbf
  {\bibinfo {volume} {57}},\ \bibinfo {pages} {439} (\bibinfo {year}
  {1983})}\BibitemShut {NoStop}%
\bibitem [{\citenamefont {Il'in}\ \emph {et~al.}(1987)\citenamefont {Il'in},
  \citenamefont {Karpov},\ and\ \citenamefont {Parshin}}]{Il-In1987-by}%
  \BibitemOpen
  \bibfield  {author} {\bibinfo {author} {\bibfnamefont {M.~A.}\ \bibnamefont
  {Il'in}}, \bibinfo {author} {\bibfnamefont {V.~G.}\ \bibnamefont {Karpov}},\
  and\ \bibinfo {author} {\bibfnamefont {D.~A.}\ \bibnamefont {Parshin}},\
  }\bibfield  {title} {\bibinfo {title} {Parameters of soft atomic potentials
  in glasses},\ }\href
  {https://ui.adsabs.harvard.edu/abs/1987JETP...65..165I/abstract} {\bibfield
  {journal} {\bibinfo  {journal} {Sov. Phys., JETP}\ }\textbf {\bibinfo
  {volume} {65}},\ \bibinfo {pages} {165} (\bibinfo {year} {1987})}\BibitemShut
  {NoStop}%
\bibitem [{\citenamefont {Buchenau}\ \emph {et~al.}(1991)\citenamefont
  {Buchenau}, \citenamefont {Galperin}, \citenamefont {Gurevich},\ and\
  \citenamefont {Schober}}]{Buchenau1991-he}%
  \BibitemOpen
  \bibfield  {author} {\bibinfo {author} {\bibfnamefont {U.}~\bibnamefont
  {Buchenau}}, \bibinfo {author} {\bibfnamefont {Y.~M.}\ \bibnamefont
  {Galperin}}, \bibinfo {author} {\bibfnamefont {V.~L.}\ \bibnamefont
  {Gurevich}},\ and\ \bibinfo {author} {\bibfnamefont {H.~R.}\ \bibnamefont
  {Schober}},\ }\bibfield  {title} {\bibinfo {title} {Anharmonic potentials and
  vibrational localization in glasses},\ }\href
  {https://doi.org/10.1103/physrevb.43.5039} {\bibfield  {journal} {\bibinfo
  {journal} {Phys. Rev. B}\ }\textbf {\bibinfo {volume} {43}},\
  \bibinfo {pages} {5039} (\bibinfo {year} {1991})}\BibitemShut {NoStop}%
\bibitem [{\citenamefont {Ji}\ \emph {et~al.}(2022)\citenamefont {Ji},
  \citenamefont {de~Geus}, \citenamefont {Agoritsas},\ and\ \citenamefont
  {Wyart}}]{Ji2022-os}%
  \BibitemOpen
  \bibfield  {author} {\bibinfo {author} {\bibfnamefont {W.}~\bibnamefont
  {Ji}}, \bibinfo {author} {\bibfnamefont {T.~W.~J.}\ \bibnamefont {de~Geus}},
  \bibinfo {author} {\bibfnamefont {E.}~\bibnamefont {Agoritsas}},\ and\
  \bibinfo {author} {\bibfnamefont {M.}~\bibnamefont {Wyart}},\ }\bibfield
  {title} {\bibinfo {title} {Mean-field description for the architecture of
  low-energy excitations in glasses},\ }\href
  {https://doi.org/10.1103/PhysRevE.105.044601} {\bibfield  {journal} {\bibinfo
   {journal} {Phys. Rev. E}\ }\textbf {\bibinfo {volume} {105}},\ \bibinfo
  {pages} {044601} (\bibinfo {year} {2022})}\BibitemShut {NoStop}%
\bibitem [{\citenamefont {Ji}\ \emph {et~al.}(2020)\citenamefont {Ji},
  \citenamefont {de~Geus}, \citenamefont {Popović}, \citenamefont
  {Agoritsas},\ and\ \citenamefont {Wyart}}]{Ji2020-qv}%
  \BibitemOpen
  \bibfield  {author} {\bibinfo {author} {\bibfnamefont {W.}~\bibnamefont
  {Ji}}, \bibinfo {author} {\bibfnamefont {T.~W.~J.}\ \bibnamefont {de~Geus}},
  \bibinfo {author} {\bibfnamefont {M.}~\bibnamefont {Popović}}, \bibinfo
  {author} {\bibfnamefont {E.}~\bibnamefont {Agoritsas}},\ and\ \bibinfo
  {author} {\bibfnamefont {M.}~\bibnamefont {Wyart}},\ }\bibfield  {title}
  {\bibinfo {title} {Thermal origin of quasilocalized excitations in glasses},\
  }\href {https://doi.org/10.1103/PhysRevE.102.062110} {\bibfield  {journal}
  {\bibinfo  {journal} {Phys. Rev. E}\ }\textbf {\bibinfo {volume} {102}},\
  \bibinfo {pages} {062110} (\bibinfo {year} {2020})}\BibitemShut {NoStop}%
\bibitem [{\citenamefont {Ji}\ \emph {et~al.}(2019)\citenamefont {Ji},
  \citenamefont {Popović}, \citenamefont {de~Geus}, \citenamefont {Lerner},\
  and\ \citenamefont {Wyart}}]{Ji2019-uj}%
  \BibitemOpen
  \bibfield  {author} {\bibinfo {author} {\bibfnamefont {W.}~\bibnamefont
  {Ji}}, \bibinfo {author} {\bibfnamefont {M.}~\bibnamefont {Popović}},
  \bibinfo {author} {\bibfnamefont {T.~W.~J.}\ \bibnamefont {de~Geus}},
  \bibinfo {author} {\bibfnamefont {E.}~\bibnamefont {Lerner}},\ and\ \bibinfo
  {author} {\bibfnamefont {M.}~\bibnamefont {Wyart}},\ }\bibfield  {title}
  {\bibinfo {title} {Theory for the density of interacting quasilocalized modes
  in amorphous solids},\ }\href {https://doi.org/10.1103/PhysRevE.99.023003}
  {\bibfield  {journal} {\bibinfo  {journal} {Phys. Rev. E}\ }\textbf
  {\bibinfo {volume} {99}},\ \bibinfo {pages} {023003} (\bibinfo {year}
  {2019})}\BibitemShut {NoStop}%
\bibitem [{\citenamefont {Karmakar}\ \emph {et~al.}(2010)\citenamefont
  {Karmakar}, \citenamefont {Lerner},\ and\ \citenamefont
  {Procaccia}}]{Karmakar2010-ft}%
  \BibitemOpen
  \bibfield  {author} {\bibinfo {author} {\bibfnamefont {S.}~\bibnamefont
  {Karmakar}}, \bibinfo {author} {\bibfnamefont {E.}~\bibnamefont {Lerner}},\
  and\ \bibinfo {author} {\bibfnamefont {I.}~\bibnamefont {Procaccia}},\
  }\bibfield  {title} {\bibinfo {title} {Statistical physics of the yielding
  transition in amorphous solids},\ }\href
  {https://doi.org/10.1103/PhysRevE.82.055103} {\bibfield  {journal} {\bibinfo
  {journal} {Phys. Rev. E}\ }\textbf {\bibinfo
  {volume} {82}},\ \bibinfo {pages} {055103} (\bibinfo {year}
  {2010})}\BibitemShut {NoStop}%
\bibitem [{\citenamefont {Hentschel}\ \emph {et~al.}(2015)\citenamefont
  {Hentschel}, \citenamefont {Jaiswal}, \citenamefont {Procaccia},\ and\
  \citenamefont {Sastry}}]{Hentschel2015-hi}%
  \BibitemOpen
  \bibfield  {author} {\bibinfo {author} {\bibfnamefont {H.~G.~E.}\
  \bibnamefont {Hentschel}}, \bibinfo {author} {\bibfnamefont {P.~K.}\
  \bibnamefont {Jaiswal}}, \bibinfo {author} {\bibfnamefont {I.}~\bibnamefont
  {Procaccia}},\ and\ \bibinfo {author} {\bibfnamefont {S.}~\bibnamefont
  {Sastry}},\ }\bibfield  {title} {\bibinfo {title} {Stochastic approach to
  plasticity and yield in amorphous solids},\ }\href
  {https://doi.org/10.1103/PhysRevE.92.062302} {\bibfield  {journal} {\bibinfo
  {journal} {Phys. Rev. E}\ }\textbf {\bibinfo
  {volume} {92}},\ \bibinfo {pages} {062302} (\bibinfo {year}
  {2015})}\BibitemShut {NoStop}%
\bibitem [{\citenamefont {Lerner}\ \emph {et~al.}(2018)\citenamefont {Lerner},
  \citenamefont {Procaccia}, \citenamefont {Rainone},\ and\ \citenamefont
  {Singh}}]{Lerner2018-ko}%
  \BibitemOpen
  \bibfield  {author} {\bibinfo {author} {\bibfnamefont {E.}~\bibnamefont
  {Lerner}}, \bibinfo {author} {\bibfnamefont {I.}~\bibnamefont {Procaccia}},
  \bibinfo {author} {\bibfnamefont {C.}~\bibnamefont {Rainone}},\ and\ \bibinfo
  {author} {\bibfnamefont {M.}~\bibnamefont {Singh}},\ }\bibfield  {title}
  {\bibinfo {title} {Protocol dependence of plasticity in ultrastable amorphous
  solids},\ }\href {https://doi.org/10.1103/physreve.98.063001} {\bibfield
  {journal} {\bibinfo  {journal} {Phys. Rev. E}\ }\textbf {\bibinfo {volume}
  {98}},\ \bibinfo {pages} {063001} (\bibinfo {year} {2018})}\BibitemShut
  {NoStop}%
\bibitem [{\citenamefont {Ruscher}\ and\ \citenamefont
  {Rottler}(2020)}]{Ruscher2020-gh}%
  \BibitemOpen
  \bibfield  {author} {\bibinfo {author} {\bibfnamefont {C.}~\bibnamefont
  {Ruscher}}\ and\ \bibinfo {author} {\bibfnamefont {J.}~\bibnamefont
  {Rottler}},\ }\bibfield  {title} {\bibinfo {title} {Residual stress
  distributions in amorphous solids from atomistic simulations},\ }\href
  {https://doi.org/10.1039/d0sm01155j} {\bibfield  {journal} {\bibinfo
  {journal} {Soft Matter}\ }\textbf {\bibinfo {volume} {16}},\ \bibinfo {pages}
  {8940} (\bibinfo {year} {2020})}\BibitemShut {NoStop}%
\bibitem [{\citenamefont {Shang}\ \emph {et~al.}(2020)\citenamefont {Shang},
  \citenamefont {Guan},\ and\ \citenamefont {Barrat}}]{Shang2020-jq}%
  \BibitemOpen
  \bibfield  {author} {\bibinfo {author} {\bibfnamefont {B.}~\bibnamefont
  {Shang}}, \bibinfo {author} {\bibfnamefont {P.}~\bibnamefont {Guan}},\ and\
  \bibinfo {author} {\bibfnamefont {J.-L.}\ \bibnamefont {Barrat}},\ }\bibfield
   {title} {\bibinfo {title} {Elastic avalanches reveal marginal behavior in
  amorphous solids},\ }\href {https://doi.org/10.1073/pnas.1915070117}
  {\bibfield  {journal} {\bibinfo  {journal} {Proc. Natl. Acad. Sci. U. S. A.}\
  }\textbf {\bibinfo {volume} {117}},\ \bibinfo {pages} {86} (\bibinfo {year}
  {2020})}\BibitemShut {NoStop}%
\bibitem [{\citenamefont {Krishnan}\ \emph {et~al.}(2022)\citenamefont
  {Krishnan}, \citenamefont {Ramola},\ and\ \citenamefont
  {Karmakar}}]{Krishnan2022-ar}%
  \BibitemOpen
  \bibfield  {author} {\bibinfo {author} {\bibfnamefont {V.~V.}\ \bibnamefont
  {Krishnan}}, \bibinfo {author} {\bibfnamefont {K.}~\bibnamefont {Ramola}},\
  and\ \bibinfo {author} {\bibfnamefont {S.}~\bibnamefont {Karmakar}},\
  }\bibfield  {title} {\bibinfo {title} {Universal non-debye low-frequency
  vibrations in sheared amorphous solids},\ }\href
  {https://doi.org/10.1039/d2sm00218c} {\bibfield  {journal} {\bibinfo
  {journal} {Soft Matter}\ }\textbf {\bibinfo {volume} {18}},\ \bibinfo {pages}
  {3395} (\bibinfo {year} {2022})}\BibitemShut {NoStop}%
\bibitem [{\citenamefont {Lin}\ \emph {et~al.}(2014)\citenamefont {Lin},
  \citenamefont {Lerner}, \citenamefont {Rosso},\ and\ \citenamefont
  {Wyart}}]{Lin2014-ov}%
  \BibitemOpen
  \bibfield  {author} {\bibinfo {author} {\bibfnamefont {J.}~\bibnamefont
  {Lin}}, \bibinfo {author} {\bibfnamefont {E.}~\bibnamefont {Lerner}},
  \bibinfo {author} {\bibfnamefont {A.}~\bibnamefont {Rosso}},\ and\ \bibinfo
  {author} {\bibfnamefont {M.}~\bibnamefont {Wyart}},\ }\bibfield  {title}
  {\bibinfo {title} {Scaling description of the yielding transition in soft
  amorphous solids at zero temperature},\ }\href
  {https://doi.org/10.1073/pnas.1406391111} {\bibfield  {journal} {\bibinfo
  {journal} {Proc. Natl. Acad. Sci. U. S. A.}\ }\textbf {\bibinfo {volume}
  {111}},\ \bibinfo {pages} {14382} (\bibinfo {year} {2014})}\BibitemShut
  {NoStop}%
\bibitem [{\citenamefont {Lin}\ and\ \citenamefont {Wyart}(2016)}]{Lin2016-jo}%
  \BibitemOpen
  \bibfield  {author} {\bibinfo {author} {\bibfnamefont {J.}~\bibnamefont
  {Lin}}\ and\ \bibinfo {author} {\bibfnamefont {M.}~\bibnamefont {Wyart}},\
  }\bibfield  {title} {\bibinfo {title} {Mean-field description of plastic flow
  in amorphous solids},\ }\href {https://doi.org/10.1103/PhysRevX.6.011005}
  {\bibfield  {journal} {\bibinfo  {journal} {Phys. Rev. X}\ }\textbf {\bibinfo
  {volume} {6}},\ \bibinfo {pages} {011005} (\bibinfo {year}
  {2016})}\BibitemShut {NoStop}%
\bibitem [{\citenamefont {Doye}\ and\ \citenamefont
  {Wales}(1997)}]{Doye1997-cz}%
  \BibitemOpen
  \bibfield  {author} {\bibinfo {author} {\bibfnamefont {J.~P.~K.}\
  \bibnamefont {Doye}}\ and\ \bibinfo {author} {\bibfnamefont {D.~J.}\
  \bibnamefont {Wales}},\ }\bibfield  {title} {\bibinfo {title} {Surveying a
  potential energy surface by eigenvector-following: Applications to global
  optimisation and the structural transformations of clusters},\ }\href
  {https://doi.org/10.1007/s004600050192} {\bibfield  {journal} {\bibinfo
  {journal} {Z. Phys. D At. Mol. Clust.}\ }\textbf {\bibinfo {volume} {40}},\
  \bibinfo {pages} {194} (\bibinfo {year} {1997})}\BibitemShut {NoStop}%
\bibitem [{\citenamefont {Malek}\ and\ \citenamefont
  {Mousseau}(2000)}]{Malek2000-mc}%
  \BibitemOpen
  \bibfield  {author} {\bibinfo {author} {\bibfnamefont {R.}~\bibnamefont
  {Malek}}\ and\ \bibinfo {author} {\bibfnamefont {N.}~\bibnamefont
  {Mousseau}},\ }\bibfield  {title} {\bibinfo {title} {Dynamics of
  lennard-jones clusters: A characterization of the activation-relaxation
  technique},\ }\href {https://doi.org/10.1103/physreve.62.7723} {\bibfield
  {journal} {\bibinfo  {journal} {Phys. Rev. E}\ }\textbf {\bibinfo {volume} {62}},\ \bibinfo
  {pages} {7723} (\bibinfo {year} {2000})}\BibitemShut {NoStop}%
\bibitem [{\citenamefont {Rodney}\ and\ \citenamefont
  {Schuh}(2009{\natexlab{a}})}]{Rodney2009-uw}%
  \BibitemOpen
  \bibfield  {author} {\bibinfo {author} {\bibfnamefont {D.}~\bibnamefont
  {Rodney}}\ and\ \bibinfo {author} {\bibfnamefont {C.}~\bibnamefont {Schuh}},\
  }\bibfield  {title} {\bibinfo {title} {Distribution of thermally activated
  plastic events in a flowing glass},\ }\href
  {https://doi.org/10.1103/PhysRevLett.102.235503} {\bibfield  {journal}
  {\bibinfo  {journal} {Phys. Rev. Lett.}\ }\textbf {\bibinfo {volume} {102}},\
  \bibinfo {pages} {235503} (\bibinfo {year} {2009}{\natexlab{a}})}\BibitemShut
  {NoStop}%
\bibitem [{\citenamefont {Rodney}\ and\ \citenamefont
  {Schuh}(2009{\natexlab{b}})}]{Rodney2009-ff}%
  \BibitemOpen
  \bibfield  {author} {\bibinfo {author} {\bibfnamefont {D.}~\bibnamefont
  {Rodney}}\ and\ \bibinfo {author} {\bibfnamefont {C.~A.}\ \bibnamefont
  {Schuh}},\ }\bibfield  {title} {\bibinfo {title} {Yield stress in metallic
  glasses: The jamming-unjamming transition studied through monte carlo
  simulations based on the activation-relaxation technique},\ }\href
  {https://doi.org/10.1103/physrevb.80.184203} {\bibfield  {journal} {\bibinfo
  {journal} {Phys. Rev. B}\ }\textbf {\bibinfo
  {volume} {80}},\ \bibinfo {pages} {184203} (\bibinfo {year}
  {2009}{\natexlab{b}})}\BibitemShut {NoStop}%
\bibitem [{\citenamefont {Kallel}\ \emph {et~al.}(2010)\citenamefont {Kallel},
  \citenamefont {Mousseau},\ and\ \citenamefont
  {Schiettekatte}}]{Kallel2010-fe}%
  \BibitemOpen
  \bibfield  {author} {\bibinfo {author} {\bibfnamefont {H.}~\bibnamefont
  {Kallel}}, \bibinfo {author} {\bibfnamefont {N.}~\bibnamefont {Mousseau}},\
  and\ \bibinfo {author} {\bibfnamefont {F.}~\bibnamefont {Schiettekatte}},\
  }\bibfield  {title} {\bibinfo {title} {Evolution of the potential-energy
  surface of amorphous silicon},\ }\href
  {https://doi.org/10.1103/PhysRevLett.105.045503} {\bibfield  {journal}
  {\bibinfo  {journal} {Phys. Rev. Lett.}\ }\textbf {\bibinfo {volume} {105}},\
  \bibinfo {pages} {045503} (\bibinfo {year} {2010})}\BibitemShut {NoStop}%
\bibitem [{\citenamefont {Koziatek}\ \emph {et~al.}(2013)\citenamefont
  {Koziatek}, \citenamefont {Barrat}, \citenamefont {Derlet},\ and\
  \citenamefont {Rodney}}]{Koziatek2013-wm}%
  \BibitemOpen
  \bibfield  {author} {\bibinfo {author} {\bibfnamefont {P.}~\bibnamefont
  {Koziatek}}, \bibinfo {author} {\bibfnamefont {J.-L.}\ \bibnamefont
  {Barrat}}, \bibinfo {author} {\bibfnamefont {P.}~\bibnamefont {Derlet}},\
  and\ \bibinfo {author} {\bibfnamefont {D.}~\bibnamefont {Rodney}},\
  }\bibfield  {title} {\bibinfo {title} {Inverse meyer-neldel behavior for
  activated processes in model glasses},\ }\href
  {https://doi.org/10.1103/physrevb.87.224105} {\bibfield  {journal} {\bibinfo
  {journal} {Phys. Rev. B}\ }\textbf {\bibinfo
  {volume} {87}},\ \bibinfo {pages} {224105} (\bibinfo {year}
  {2013})}\BibitemShut {NoStop}%
\bibitem [{\citenamefont {Swayamjyoti}\ \emph {et~al.}(2014)\citenamefont
  {Swayamjyoti}, \citenamefont {Löffler},\ and\ \citenamefont
  {Derlet}}]{Swayamjyoti2014-pw}%
  \BibitemOpen
  \bibfield  {author} {\bibinfo {author} {\bibfnamefont {S.}~\bibnamefont
  {Swayamjyoti}}, \bibinfo {author} {\bibfnamefont {J.~F.}\ \bibnamefont
  {Löffler}},\ and\ \bibinfo {author} {\bibfnamefont {P.~M.}\ \bibnamefont
  {Derlet}},\ }\bibfield  {title} {\bibinfo {title} {Local structural
  excitations in model glasses},\ }\href
  {https://doi.org/10.1103/physrevb.89.224201} {\bibfield  {journal} {\bibinfo
  {journal} {Phys. Rev. B}\ }\textbf {\bibinfo
  {volume} {89}},\ \bibinfo {pages} {224201} (\bibinfo {year}
  {2014})}\BibitemShut {NoStop}%
\bibitem [{\citenamefont {Fan}\ \emph {et~al.}(2015)\citenamefont {Fan},
  \citenamefont {Iwashita},\ and\ \citenamefont {Egami}}]{Fan2015-vi}%
  \BibitemOpen
  \bibfield  {author} {\bibinfo {author} {\bibfnamefont {Y.}~\bibnamefont
  {Fan}}, \bibinfo {author} {\bibfnamefont {T.}~\bibnamefont {Iwashita}},\ and\
  \bibinfo {author} {\bibfnamefont {T.}~\bibnamefont {Egami}},\ }\bibfield
  {title} {\bibinfo {title} {Crossover from localized to cascade relaxations in
  metallic glasses},\ }\href {https://doi.org/10.1103/PhysRevLett.115.045501}
  {\bibfield  {journal} {\bibinfo  {journal} {Phys. Rev. Lett.}\ }\textbf
  {\bibinfo {volume} {115}},\ \bibinfo {pages} {045501} (\bibinfo {year}
  {2015})}\BibitemShut {NoStop}%
\bibitem [{\citenamefont {Rodney}\ \emph {et~al.}(2011)\citenamefont {Rodney},
  \citenamefont {Tanguy},\ and\ \citenamefont {Vandembroucq}}]{Rodney2011-mo}%
  \BibitemOpen
  \bibfield  {author} {\bibinfo {author} {\bibfnamefont {D.}~\bibnamefont
  {Rodney}}, \bibinfo {author} {\bibfnamefont {A.}~\bibnamefont {Tanguy}},\
  and\ \bibinfo {author} {\bibfnamefont {D.}~\bibnamefont {Vandembroucq}},\
  }\bibfield  {title} {\bibinfo {title} {Modeling the mechanics of amorphous
  solids at different length scale and time scale},\ }\href
  {https://doi.org/10.1088/0965-0393/19/8/083001} {\bibfield  {journal}
  {\bibinfo  {journal} {Model. Simul. Mater. Sci. Eng.}\ }\textbf {\bibinfo
  {volume} {19}},\ \bibinfo {pages} {083001} (\bibinfo {year}
  {2011})}\BibitemShut {NoStop}%
\bibitem [{\citenamefont {Doliwa}\ and\ \citenamefont
  {Heuer}(2003)}]{Doliwa2003-ow}%
  \BibitemOpen
  \bibfield  {author} {\bibinfo {author} {\bibfnamefont {B.}~\bibnamefont
  {Doliwa}}\ and\ \bibinfo {author} {\bibfnamefont {A.}~\bibnamefont {Heuer}},\
  }\bibfield  {title} {\bibinfo {title} {Energy barriers and activated dynamics
  in a supercooled lennard-jones liquid},\ }\href
  {https://doi.org/10.1103/PhysRevE.67.031506} {\bibfield  {journal} {\bibinfo
  {journal} {Phys. Rev. E}\ }\textbf {\bibinfo
  {volume} {67}},\ \bibinfo {pages} {031506} (\bibinfo {year}
  {2003})}\BibitemShut {NoStop}%
\bibitem [{\citenamefont {Reinisch}\ and\ \citenamefont
  {Heuer}(2005)}]{Reinisch2005-me}%
  \BibitemOpen
  \bibfield  {author} {\bibinfo {author} {\bibfnamefont {J.}~\bibnamefont
  {Reinisch}}\ and\ \bibinfo {author} {\bibfnamefont {A.}~\bibnamefont
  {Heuer}},\ }\bibfield  {title} {\bibinfo {title} {What is moving in silica at
  1 {K}? a computer study of the low-temperature anomalies},\ }\href
  {https://doi.org/10.1103/PhysRevLett.95.155502} {\bibfield  {journal}
  {\bibinfo  {journal} {Phys. Rev. Lett.}\ }\textbf {\bibinfo {volume} {95}},\
  \bibinfo {pages} {155502} (\bibinfo {year} {2005})}\BibitemShut {NoStop}%
\bibitem [{\citenamefont {Hamdan}\ \emph {et~al.}(2014)\citenamefont {Hamdan},
  \citenamefont {Trinastic},\ and\ \citenamefont {Cheng}}]{Hamdan2014-lj}%
  \BibitemOpen
  \bibfield  {author} {\bibinfo {author} {\bibfnamefont {R.}~\bibnamefont
  {Hamdan}}, \bibinfo {author} {\bibfnamefont {J.~P.}\ \bibnamefont
  {Trinastic}},\ and\ \bibinfo {author} {\bibfnamefont {H.~P.}\ \bibnamefont
  {Cheng}},\ }\bibfield  {title} {\bibinfo {title} {Molecular dynamics study of
  the mechanical loss in amorphous pure and doped silica},\ }\href
  {https://doi.org/10.1063/1.4890958} {\bibfield  {journal} {\bibinfo
  {journal} {J. Chem. Phys.}\ }\textbf {\bibinfo {volume} {141}},\ \bibinfo
  {pages} {054501} (\bibinfo {year} {2014})}\BibitemShut {NoStop}%
\bibitem [{\citenamefont {Damart}\ and\ \citenamefont
  {Rodney}(2018)}]{Damart2018-bp}%
  \BibitemOpen
  \bibfield  {author} {\bibinfo {author} {\bibfnamefont {T.}~\bibnamefont
  {Damart}}\ and\ \bibinfo {author} {\bibfnamefont {D.}~\bibnamefont
  {Rodney}},\ }\bibfield  {title} {\bibinfo {title} {Atomistic study of
  two-level systems in amorphous silica},\ }\href
  {https://doi.org/10.1103/physrevb.97.014201} {\bibfield  {journal} {\bibinfo
  {journal} {Phys. Rev. B}\ }\textbf {\bibinfo {volume} {97}},\ \bibinfo
  {pages} {014201} (\bibinfo {year} {2018})}\BibitemShut {NoStop}%
\bibitem [{\citenamefont {Khomenko}\ \emph {et~al.}(2020)\citenamefont
  {Khomenko}, \citenamefont {Scalliet}, \citenamefont {Berthier}, \citenamefont
  {Reichman},\ and\ \citenamefont {Zamponi}}]{Khomenko2020-gk}%
  \BibitemOpen
  \bibfield  {author} {\bibinfo {author} {\bibfnamefont {D.}~\bibnamefont
  {Khomenko}}, \bibinfo {author} {\bibfnamefont {C.}~\bibnamefont {Scalliet}},
  \bibinfo {author} {\bibfnamefont {L.}~\bibnamefont {Berthier}}, \bibinfo
  {author} {\bibfnamefont {D.~R.}\ \bibnamefont {Reichman}},\ and\ \bibinfo
  {author} {\bibfnamefont {F.}~\bibnamefont {Zamponi}},\ }\bibfield  {title}
  {\bibinfo {title} {Depletion of two-level systems in ultrastable
  computer-generated glasses},\ }\href
  {https://doi.org/10.1103/PhysRevLett.124.225901} {\bibfield  {journal}
  {\bibinfo  {journal} {Phys. Rev. Lett.}\ }\textbf {\bibinfo {volume} {124}},\
  \bibinfo {pages} {225901} (\bibinfo {year} {2020})}\BibitemShut {NoStop}%
\bibitem [{\citenamefont {Khomenko}\ \emph {et~al.}(2021)\citenamefont
  {Khomenko}, \citenamefont {Reichman},\ and\ \citenamefont
  {Zamponi}}]{Khomenko2021-js}%
  \BibitemOpen
  \bibfield  {author} {\bibinfo {author} {\bibfnamefont {D.}~\bibnamefont
  {Khomenko}}, \bibinfo {author} {\bibfnamefont {D.~R.}\ \bibnamefont
  {Reichman}},\ and\ \bibinfo {author} {\bibfnamefont {F.}~\bibnamefont
  {Zamponi}},\ }\bibfield  {title} {\bibinfo {title} {Relationship between
  two-level systems and quasilocalized normal modes in glasses},\ }\href
  {https://doi.org/10.1103/physrevmaterials.5.055602} {\bibfield  {journal}
  {\bibinfo  {journal} {Phys. Rev. Mater.}\ }\textbf {\bibinfo {volume} {5}},\
  \bibinfo {pages} {055602} (\bibinfo {year} {2021})}\BibitemShut {NoStop}%
\bibitem [{\citenamefont {Mocanu}\ \emph {et~al.}(2023)\citenamefont {Mocanu},
  \citenamefont {Berthier}, \citenamefont {Ciarella}, \citenamefont {Khomenko},
  \citenamefont {Reichman}, \citenamefont {Scalliet},\ and\ \citenamefont
  {Zamponi}}]{Mocanu2023-kv}%
  \BibitemOpen
  \bibfield  {author} {\bibinfo {author} {\bibfnamefont {F.~C.}\ \bibnamefont
  {Mocanu}}, \bibinfo {author} {\bibfnamefont {L.}~\bibnamefont {Berthier}},
  \bibinfo {author} {\bibfnamefont {S.}~\bibnamefont {Ciarella}}, \bibinfo
  {author} {\bibfnamefont {D.}~\bibnamefont {Khomenko}}, \bibinfo {author}
  {\bibfnamefont {D.~R.}\ \bibnamefont {Reichman}}, \bibinfo {author}
  {\bibfnamefont {C.}~\bibnamefont {Scalliet}},\ and\ \bibinfo {author}
  {\bibfnamefont {F.}~\bibnamefont {Zamponi}},\ }\bibfield  {title} {\bibinfo
  {title} {Microscopic observation of two-level systems in a metallic glass
  model},\ }\href {https://doi.org/10.1063/5.0128820} {\bibfield  {journal}
  {\bibinfo  {journal} {J. Chem. Phys.}\ }\textbf {\bibinfo {volume} {158}},\
  \bibinfo {pages} {014501} (\bibinfo {year} {2023})}\BibitemShut {NoStop}%
\bibitem [{\citenamefont {Ciarella}\ \emph {et~al.}(2023)\citenamefont
  {Ciarella}, \citenamefont {Khomenko}, \citenamefont {Berthier}, \citenamefont
  {Mocanu}, \citenamefont {Reichman}, \citenamefont {Scalliet},\ and\
  \citenamefont {Zamponi}}]{Ciarella2023-zf}%
  \BibitemOpen
  \bibfield  {author} {\bibinfo {author} {\bibfnamefont {S.}~\bibnamefont
  {Ciarella}}, \bibinfo {author} {\bibfnamefont {D.}~\bibnamefont {Khomenko}},
  \bibinfo {author} {\bibfnamefont {L.}~\bibnamefont {Berthier}}, \bibinfo
  {author} {\bibfnamefont {F.~C.}\ \bibnamefont {Mocanu}}, \bibinfo {author}
  {\bibfnamefont {D.~R.}\ \bibnamefont {Reichman}}, \bibinfo {author}
  {\bibfnamefont {C.}~\bibnamefont {Scalliet}},\ and\ \bibinfo {author}
  {\bibfnamefont {F.}~\bibnamefont {Zamponi}},\ }\bibfield  {title} {\bibinfo
  {title} {Finding defects in glasses through machine learning},\ }\href
  {https://doi.org/10.1038/s41467-023-39948-7} {\bibfield  {journal} {\bibinfo
  {journal} {Nat. Commun.}\ }\textbf {\bibinfo {volume} {14}},\ \bibinfo
  {pages} {4229} (\bibinfo {year} {2023})}\BibitemShut {NoStop}%
\bibitem [{\citenamefont {Pica~Ciamarra}\ \emph {et~al.}(2024)\citenamefont
  {Pica~Ciamarra}, \citenamefont {Ji},\ and\ \citenamefont
  {Wyart}}]{Pica-Ciamarra2024-xk}%
  \BibitemOpen
  \bibfield  {author} {\bibinfo {author} {\bibfnamefont {M.}~\bibnamefont
  {Pica~Ciamarra}}, \bibinfo {author} {\bibfnamefont {W.}~\bibnamefont {Ji}},\
  and\ \bibinfo {author} {\bibfnamefont {M.}~\bibnamefont {Wyart}},\ }\bibfield
   {title} {\bibinfo {title} {Local vs. cooperative: Unraveling glass
  transition mechanisms with {SEER}},\ }\href
  {https://doi.org/10.1073/pnas.2400611121} {\bibfield  {journal} {\bibinfo
  {journal} {Proc. Natl. Acad. Sci. U. S. A.}\ }\textbf {\bibinfo {volume}
  {121}},\ \bibinfo {pages} {e2400611121} (\bibinfo {year} {2024})}\BibitemShut
  {NoStop}%
\bibitem [{\citenamefont {Ji}\ \emph {et~al.}(2025)\citenamefont {Ji},
  \citenamefont {Pica~Ciamarra},\ and\ \citenamefont {Wyart}}]{Ji2025-xk}%
  \BibitemOpen
  \bibfield  {author} {\bibinfo {author} {\bibfnamefont {W.}~\bibnamefont
  {Ji}}, \bibinfo {author} {\bibfnamefont {M.}~\bibnamefont {Pica~Ciamarra}},\
  and\ \bibinfo {author} {\bibfnamefont {M.}~\bibnamefont {Wyart}},\ }\bibfield
   {title} {\bibinfo {title} {The role of excitations in supercooled liquids:
  Density, geometry, and relaxation dynamics},\ }\href
  {https://doi.org/10.1073/pnas.2416800122} {\bibfield  {journal} {\bibinfo
  {journal} {Proc. Natl. Acad. Sci. U. S. A.}\ }\textbf {\bibinfo {volume}
  {122}},\ \bibinfo {pages} {e2416800122} (\bibinfo {year} {2025})}\BibitemShut
  {NoStop}%
\bibitem [{\citenamefont {Ghanem}\ \emph {et~al.}(2026)\citenamefont {Ghanem},
  \citenamefont {Jana}, \citenamefont {Nöhring}, \citenamefont {Moseler},\
  and\ \citenamefont {Pastewka}}]{Ghanem2026-ui}%
  \BibitemOpen
  \bibfield  {author} {\bibinfo {author} {\bibfnamefont {I.}~\bibnamefont
  {Ghanem}}, \bibinfo {author} {\bibfnamefont {R.}~\bibnamefont {Jana}},
  \bibinfo {author} {\bibfnamefont {W.~G.}\ \bibnamefont {Nöhring}}, \bibinfo
  {author} {\bibfnamefont {M.}~\bibnamefont {Moseler}},\ and\ \bibinfo {author}
  {\bibfnamefont {L.}~\bibnamefont {Pastewka}},\ }\bibfield  {title} {\bibinfo
  {title} {Universal energy barrier for plastic flow in a network glass},\
  }\bibfield  {journal} {\bibinfo  {journal} {arXiv [cond-mat.dis-nn]}\ }\href
  {https://doi.org/10.48550/arXiv.2610.02454} {10.48550/arXiv.2610.02454}
  (\bibinfo {year} {2026})\BibitemShut {NoStop}%
\bibitem [{\citenamefont {Henkelman}\ \emph {et~al.}(2000)\citenamefont
  {Henkelman}, \citenamefont {Uberuaga},\ and\ \citenamefont
  {Jónsson}}]{Henkelman2000-hr}%
  \BibitemOpen
  \bibfield  {author} {\bibinfo {author} {\bibfnamefont {G.}~\bibnamefont
  {Henkelman}}, \bibinfo {author} {\bibfnamefont {B.~P.}\ \bibnamefont
  {Uberuaga}},\ and\ \bibinfo {author} {\bibfnamefont {H.}~\bibnamefont
  {Jónsson}},\ }\bibfield  {title} {\bibinfo {title} {A climbing image nudged
  elastic band method for finding saddle points and minimum energy paths},\
  }\href {https://doi.org/10.1063/1.1329672} {\bibfield  {journal} {\bibinfo
  {journal} {J. Chem. Phys.}\ }\textbf {\bibinfo {volume} {113}},\ \bibinfo
  {pages} {9901} (\bibinfo {year} {2000})}\BibitemShut {NoStop}%
\bibitem [{\citenamefont {Olsen}\ \emph {et~al.}(2004)\citenamefont {Olsen},
  \citenamefont {Kroes}, \citenamefont {Henkelman}, \citenamefont
  {Arnaldsson},\ and\ \citenamefont {Jónsson}}]{Olsen2004-px}%
  \BibitemOpen
  \bibfield  {author} {\bibinfo {author} {\bibfnamefont {R.~A.}\ \bibnamefont
  {Olsen}}, \bibinfo {author} {\bibfnamefont {G.~J.}\ \bibnamefont {Kroes}},
  \bibinfo {author} {\bibfnamefont {G.}~\bibnamefont {Henkelman}}, \bibinfo
  {author} {\bibfnamefont {A.}~\bibnamefont {Arnaldsson}},\ and\ \bibinfo
  {author} {\bibfnamefont {H.}~\bibnamefont {Jónsson}},\ }\bibfield  {title}
  {\bibinfo {title} {Comparison of methods for finding saddle points without
  knowledge of the final states},\ }\href {https://doi.org/10.1063/1.1809574}
  {\bibfield  {journal} {\bibinfo  {journal} {J. Chem. Phys.}\ }\textbf
  {\bibinfo {volume} {121}},\ \bibinfo {pages} {9776} (\bibinfo {year}
  {2004})}\BibitemShut {NoStop}%
\bibitem [{\citenamefont {Mizuno}\ \emph {et~al.}(2020)\citenamefont {Mizuno},
  \citenamefont {Shimada},\ and\ \citenamefont {Ikeda}}]{Mizuno2020-uq}%
  \BibitemOpen
  \bibfield  {author} {\bibinfo {author} {\bibfnamefont {H.}~\bibnamefont
  {Mizuno}}, \bibinfo {author} {\bibfnamefont {M.}~\bibnamefont {Shimada}},\
  and\ \bibinfo {author} {\bibfnamefont {A.}~\bibnamefont {Ikeda}},\ }\bibfield
   {title} {\bibinfo {title} {Anharmonic properties of vibrational excitations
  in amorphous solids},\ }\href
  {https://doi.org/10.1103/physrevresearch.2.013215} {\bibfield  {journal}
  {\bibinfo  {journal} {Phys. Rev. Res.}\ }\textbf {\bibinfo {volume} {2}},\
  \bibinfo {pages} {013215} (\bibinfo {year} {2020})}\BibitemShut {NoStop}%
\bibitem [{Note1()}]{Note1}%
  \BibitemOpen
  \bibinfo {note} {Reference~\cite {Mizuno2020-uq} studied anharmonic glassy
  landscape properties along generalized directions corresponding to soft,
  quasi-localized eigenmodes, detecting and quantifying barriers along these
  directions. While these barriers, denoted by $\Delta \Phi _{\protect \rm
  barrier}$ therein, are not associated with actual landscape saddles and thus
  differ from $\Delta {U}$ discussed in this work, the data presented in
  Figs.~3 and~5 of~\cite {Mizuno2020-uq} are scaling-wise consistent with
  Eqs.~\protect \eqref {eq:Delta_U_f_c_relations} and~\protect \eqref
  {eq:Delta_U_w4}, respectively.}\BibitemShut {Stop}%
\bibitem [{\citenamefont {Nagasawa}\ \emph {et~al.}(2026)\citenamefont
  {Nagasawa}, \citenamefont {Martens}, \citenamefont {Barrat},\ and\
  \citenamefont {Ozawa}}]{Nagasawa2026-yr}%
  \BibitemOpen
  \bibfield  {author} {\bibinfo {author} {\bibfnamefont {T.}~\bibnamefont
  {Nagasawa}}, \bibinfo {author} {\bibfnamefont {K.}~\bibnamefont {Martens}},
  \bibinfo {author} {\bibfnamefont {J.-L.}\ \bibnamefont {Barrat}},\ and\
  \bibinfo {author} {\bibfnamefont {M.}~\bibnamefont {Ozawa}},\ }\bibfield
  {title} {\bibinfo {title} {Response of a model glass to athermal quasistatic
  pinching},\ }\bibfield  {journal} {\bibinfo  {journal} {arXiv
  [cond-mat.soft]}\ }\href {https://doi.org/10.48550/arXiv.2609.02692}
  {10.48550/arXiv.2609.02692} (\bibinfo {year} {2026})\BibitemShut {NoStop}%
\bibitem [{\citenamefont {Birks}\ \emph {et~al.}(2026)\citenamefont {Birks},
  \citenamefont {Ghanem}, \citenamefont {Pastewka}, \citenamefont {Kermode},\
  and\ \citenamefont {Buze}}]{Birks2026-wy}%
  \BibitemOpen
  \bibfield  {author} {\bibinfo {author} {\bibfnamefont {F.}~\bibnamefont
  {Birks}}, \bibinfo {author} {\bibfnamefont {I.}~\bibnamefont {Ghanem}},
  \bibinfo {author} {\bibfnamefont {L.}~\bibnamefont {Pastewka}}, \bibinfo
  {author} {\bibfnamefont {J.}~\bibnamefont {Kermode}},\ and\ \bibinfo {author}
  {\bibfnamefont {M.}~\bibnamefont {Buze}},\ }\bibfield  {title} {\bibinfo
  {title} {Resolving structural avalanches in amorphous carbon with arclength
  continuation},\ }\href {https://doi.org/10.1103/6n5m-rxc1} {\bibfield
  {journal} {\bibinfo  {journal} {Phys. Rev. Lett.}\ }\textbf {\bibinfo
  {volume} {136}},\ \bibinfo {pages} {206101} (\bibinfo {year}
  {2026})}\BibitemShut {NoStop}%
\bibitem [{\citenamefont {Richard}\ \emph {et~al.}(2023)\citenamefont
  {Richard}, \citenamefont {Kapteijns},\ and\ \citenamefont
  {Lerner}}]{Richard2023-bo}%
  \BibitemOpen
  \bibfield  {author} {\bibinfo {author} {\bibfnamefont {D.}~\bibnamefont
  {Richard}}, \bibinfo {author} {\bibfnamefont {G.}~\bibnamefont {Kapteijns}},\
  and\ \bibinfo {author} {\bibfnamefont {E.}~\bibnamefont {Lerner}},\
  }\bibfield  {title} {\bibinfo {title} {Detecting low-energy quasilocalized
  excitations in computer glasses},\ }\href
  {https://doi.org/10.1103/PhysRevE.108.044124} {\bibfield  {journal} {\bibinfo
   {journal} {Phys. Rev. E}\ }\textbf {\bibinfo {volume} {108}},\ \bibinfo
  {pages} {044124} (\bibinfo {year} {2023})}\BibitemShut {NoStop}%
\bibitem [{\citenamefont {Bitzek}\ \emph {et~al.}(2006)\citenamefont {Bitzek},
  \citenamefont {Koskinen}, \citenamefont {Gähler}, \citenamefont {Moseler},\
  and\ \citenamefont {Gumbsch}}]{Bitzek2006-ht}%
  \BibitemOpen
  \bibfield  {author} {\bibinfo {author} {\bibfnamefont {E.}~\bibnamefont
  {Bitzek}}, \bibinfo {author} {\bibfnamefont {P.}~\bibnamefont {Koskinen}},
  \bibinfo {author} {\bibfnamefont {F.}~\bibnamefont {Gähler}}, \bibinfo
  {author} {\bibfnamefont {M.}~\bibnamefont {Moseler}},\ and\ \bibinfo {author}
  {\bibfnamefont {P.}~\bibnamefont {Gumbsch}},\ }\bibfield  {title} {\bibinfo
  {title} {Structural relaxation made simple},\ }\href
  {https://doi.org/10.1103/PhysRevLett.97.170201} {\bibfield  {journal}
  {\bibinfo  {journal} {Phys. Rev. Lett.}\ }\textbf {\bibinfo {volume} {97}},\
  \bibinfo {pages} {170201} (\bibinfo {year} {2006})}\BibitemShut {NoStop}%
\end{thebibliography}

%

\end{document}